\documentstyle[aps,psfig]{revtex}

\begin{document}

\title{Study of the spatial and temporal
coherence of
high order harmonics}
\author{Pascal Sali\`eres$^1$,  Anne L'Huillier$^2$,
 Philippe Antoine$^{3}$
and Maciej Lewenstein$^{1}$}
\address{ 
(1) Commissariat \`a l'Energie Atomique, DSM/DRECAM/SPAM,
Centre d'Etudes de Saclay, 91191 Gif--sur--Yvette, France
}
\address{
(2) Department of Physics, Lund Institute of Technology, 
S-221 00 Lund, Sweden}
\address{
(3) Laboratoire de Physique Atomique et Mol\'eculaire,
 Universit\'e Catholique de Louvain,
 chemin du cyclotron, 2 B-1348 Louvain-la-Neuve, Belgium}
\maketitle

\begin{abstract}
We apply the theory of high-order harmonic
generation by low-frequency laser fields in the strong field approximation
to the study of the spatial and temporal coherence properties of the harmonics.
We discuss the role of  dynamically induced 
phases of the atomic polarization in determining the optimal
phase matching conditions and angular distributions of harmonics.
We demonstrate that the phase matching and the spatial coherence
can be controlled by changing the focusing 
parameters of the fundamental laser beam.
Then we present a detailed study of the temporal and spectral properties
of harmonics. We discuss how the focusing conditions influence 
the individual harmonic spectra and time profiles, and how the intensity dependence of the dynamically induced phase leads to a chirp of the harmonic frequency. This phase modulation can be used to control the temporal and spectral properties of the harmonic radiation. Temporally, the harmonic chirped pulse can be recompressed to very small durations. 
Spectrally, chirping of the fundamental beam may be employed to compensate for
the dynamically induced chirp and to control the individual 
harmonic spectrum. Finally, we discuss the short pulse effects, in particular nonadiabatic phenomena and the possibility of generating attosecond pulses.

\end{abstract}
\tableofcontents
\section{Introduction}

\subsection{Short history of high harmonic generation}

During the recent years high-order harmonic generation (HG)
has become one of the
major topics of super intense laser-atom physics. Generally speaking,
high harmonics are generated
when a short, intense laser pulse interacts with matter.  Although HG has been
in the course of recent years mainly studied in atomic gases, it has also been investigated in ions, molecules, atom clusters and solids.  Apart from its
 fascinating fundamental aspects, HG has become one of the most promising ways of producing short-pulse coherent radiation in the XUV range.
HG has already been a subject
of several review articles \cite{atom,Miyazaki,Protopapas}. One can 
point out the following 
milestones in the short history of this subject:

\begin{itemize}

\item {\it First observations} High harmonic generation is an entirely 
nonlinear and non-perturbative process. The spectrum of high harmonics
is characterized by a fall-off for the few low order harmonics,
followed by an extended {\it  plateau}, and by a rapid {\it cut-off}. 
The first experimental observations of the plateau were accomplished by
 \cite{McPherson} and \cite{Ferray} in the end of 1980's.

\item{\it Plateau extention} Most of the early work has concentrated on the 
extension of the plateau, i.e. generation of harmonics of higher frequency and shorter wavelength \cite{harg,annephil}. By focusing short-pulse terawatt lasers in rare gas jets, wavelengths as short as 7.4 nm (143rd harmonic of a 1053 nm Nd-Glass laser \cite{Perry}), 6.7 nm (37th harmonic of a 248 nm KrF laser \cite{preston}) and 4.7 nm (169th harmonic of an 800 nm Ti-Sapphire laser \cite{Chang}) have been obtained. Very recently, with ultra
short intense infrared pulses, it has become possible to generate XUV radiation extending to the water window (below the carbon K-edge at 4.4nm) \cite{Spielmann,Changprl}.

\item{\it Simple man's theory}  A breakthrough in the theoretical understanding of HG process in low frequency laser fields
was initiated by \cite{Krause} who have shown
that the cut-off position in the harmonic spectrum follows the universal law
$I_p+3U_p$, where $I_p$ is the ionization potential, whereas $U_p=e^2{\cal E}^2/4m\omega^2$, is the ponderomotive potential, i.e. the mean kinetic energy acquired by an electron oscillating in the laser field. Here $e$ is the electron charge, $m$ is its mass, and 
$\cal E$ and $\omega$  are the laser electric field and its frequency, respectively. Pretty soon an explanation of this universal fact in the framework of ``simple man's theory'' was found \cite{Kulsilap,Corkum}. According to this theory, harmonic generation occurs in the following manner: first the electron tunnels  
out from the nucleus through the 
Coulomb energy barrier modified by the presence of the (relatively slowly varying) electric  field of the laser. It then undergoes oscillations in the field, during which the influence of the Coulomb force from the nucleus is practically negligible. Finally, if the electron comes back to the vicinity of the nucleus, it may recombine back to the ground state, thus producing a photon of energy $I_p$ plus the kinetic energy acquired during the oscillatory motion.
According to classical mechanics, the maximal kinetic energy that the electron can gain is indeed $\simeq 3U_p$. A fully quantum mechanical theory, that 
is based on
strong field approximation and that recovers the ``simple man's theory'', was formulated soon after \cite{PRL,opus}.

\item{\it Ellipticity studies} The ``simple man's theory'' leads to the immediate
consequence that harmonic generation in elliptically polarized 
fields should be strongly suppressed, since the electron  released from the nucleus in such fields practically never comes back, and thus cannot recombine \cite{Corkum,Corkum2}. Several groups have demonstrated this effect \cite{Budil,Dietrich,Liang}, and have since then performed systematic experimental \cite{Burnett,Weihe,Antoine3,weihe1,schulze} 
and theoretical \cite{becker,ellip,compbeck} studies of the polarization properties of harmonics generated by elliptically polarized fields.

\item{\it Optimization and control} Progress in experimental techniques and theoretical understanding has stimulated numerous studies of optimization 
and control of HG depending on various parameters of the laser and the active medium. These studies involved among others:
\begin{itemize}
\item{\it Optimization of laser parameters} These studies concern for
instance laser polarization (discussed above), pulse duration or
 wavelength dependence \cite{balc92,kond93,Christov,balthese,these}. Although typically infrared (Nd-Glass or Ti-Sapphire) lasers are used, generation by shorter wavelength 
intense KrF lasers is also very efficient (\cite{preston}, for theory see
\cite{sanp}). 
\item{\it Generation by multicolored fields} Harmonic generation in 
combined laser fields of two frequencies was studied in the context of 
(i) enhancement of conversion efficiency (for theory see \cite{Hannover,proto2,Telnov,kondo,Perry2}, for experiment cf. \cite{Watanabe,paulus}),
(ii) access to new frequencies and tunability, if one of the fields is
tunable (for theory see \cite{Mette}, for experiments \cite{Eichmann,exp}),
and (iii) the control of HG process in general \cite{Ivanov}.
\item{\it Optimization of the generating medium} First of all, optimization with respect to atomic gases was studied
\cite{balc92,balc93}. Other active
media apart from noble gases have been used to generate harmonics:
ions (these involve ionized noble gas atoms \cite{sarak91,suedois,kond94,preston},
and alkaline ions \cite{akiyama,wahlstrom}), molecular gases \cite{chin,fraser,lynga}, atomic clusters \cite{Ditmire} etc. At this point
it is worth adding that harmonic generation from solid targets and laser
induced plasmas
has also been intensively studied in the recent years
(for theory see \cite{Gibbon,pukhov,lichter,roso1,roso2}, for experiments cf.
\cite{co2,wahl2,vonder,norrey}).
\item{\it Optimization and characteristics 
of spatial and temporal properties}  Those studies concern spatial, temporal, and spectral properties of harmonic radiation, and in particular their  coherence properties \cite{these}; they are closely related to the subject of this review and will be discussed separately below. Other examples
 of such studies involve spatial control of HG using
spatially dependent ellipticity \cite{Mercer}, control of phase matching conditions  for low \cite{Meyer} and high harmonics \cite{prlphase},
role of ionisation and defocusing effects \cite{altucci,Miyazaki,ditmi95}.
\item{\it Other control schemes}. Other control schemes of harmonic generation
have been proposed that involve for example the coherent superposition of atomic states
\cite{super,sanpml}
\end{itemize}

\item{\it Applications} High harmonics provide a very promising  source of coherent XUV radiation, with numerous applications in various areas of physics. In particular, applications in atomic physics are reviewed
 by \cite{balc95,application}. Harmonics have already been used for solid state spectroscopy \cite{Haight},
and plasma diagnostics \cite{Theobald}. Further applications that employ directly coherence properties of harmonics will be discussed in this review.

\item{\it Attosecond physics} Future applications of high harmonics will
presumably involve attosecond physics, i.e. the physics of generation,
control, detection and application of sub-femtosecond laser pulses. Two types of proposals how to reach
the
subfemtosecond limit have
been put forward over the last few years:  those that
rely on phase locking between
consecutive
harmonics \cite{Farkas,Harris,Antoine,Corkum2,Ivanov,wahl}, and those that concern single
harmonics \cite{kenpriv,these}. 

\end{itemize}

\subsection{Spatio-temporal characteristics of high harmonics}

\subsubsection{Experiments}

Both from the fundamental and practical  points of view it is
 very important  to know and understand the spatial and temporal coherence properties of high harmonics.
 Informations on the spatial coherence of the beam and on its focusability, its spectrum and time profile  are of direct interest for applications. But
 they also help in understanding the physics of the process, since  
there are many possible causes for distortion of the spatial and temporal profiles, and their interpretation implies a rather refined and deep study
of the problem.

The spatial distribution of the harmonic emission has been investigated by several groups in various experimental conditions. Peatross and Meyerhofer \cite{peatross} used a 1 $\mu$m 1 ps Nd-Glass laser loosely focused (f/70) into a very diluted gaseous media (1 Torr) in order to get rid of distortions induced by phase matching and propagation in the medium. The far-field distributions of the harmonics (11 to 41) generated in heavy rare gases were found to be quite distorted, with pedestals surrounding a narrow central peak. These wings were attributed to the rapid variation of the harmonic dipole phase with the laser intensity. 

Tisch {\it et al.} \cite{tisch} studied high-order harmonics (71 to 111) generated by a similar laser, focused (f/50) in 10 Torr of helium. Complex spatial distributions are found for harmonics in the plateau region of the spectrum. However, in the cutoff, the measured angular distributions narrow to approximately that predicted by lowest-order perturbation theory. The broad distributions with numerous substructures observed in the plateau are attributed to the influence of ionization, and in particular of the free electrons, on phase matching. 

The influence of ionization on spatial profiles has also been investigated experimentally by L'Huillier and Balcou \cite{annebal} for low-order harmonics in xenon, and, more recently, by Wahlstr\"om {\it et al.}
\cite{wahlstrom} for harmonics generated by rare-gas-like ions. Generally speaking, ionization induces a significant distortion of the harmonic profiles, thus complicating their interpretation. In a recent Letter \cite{salieres}, we presented results of an experimental study of spatial profiles of harmonics generated by a 140 fs Cr:LiSrAlF6 (Cr:LiSAF) laser system. Thanks to this very short pulse duration, it was possible to expose the medium to high intensities while keeping a weak degree of ionization. Under certain conditions, the resulting harmonic profiles were found to be very smooth, Gaussian to near flat-top, without substructure.  

In Ref. \cite{salier2}, we present systematic experimental studies of harmonic angular distributions, investigating the influence of different parameters, such as laser intensity, nonlinear order, nature of the gas and position of the laser focus relative to the generating medium. We show that, when the laser is focused before the atomic medium, harmonics with regular spatial profiles can be generated with reasonable conversion efficiency. Their divergence does not depend directly on the nonlinear order, the intensity or even the nature of the generating gas, but rather on the region of the spectrum the considered harmonic belongs to, which is determined by the combination of the three preceding elements. When the focus is drawn closer to the medium, the distributions get increasingly distorted, becoming annular with a significant divergence for a focus right within or after the jet. 

A first endeavour to measure the degree of spatial coherence of the harmonic radiation has been recently made by Ditmire {\it et al.} \cite{ditmire} with a Young two-slit experiment. They investigated how
the coherence between two points chosen to be located symmetrically 
relative to the propagation axis depends on the degree of ionization of the medium. Finally, very recently  far-field interference pattern created by overlaping in space two beams of the 13th harmonic, generated independently at different places in a xenon gas jet was observed \cite{wahl,zerne}.

The experimental studies of temporal and spectral properties of high-order harmonics have also been carried out in different experimental conditions. Temporal profiles of low harmonics generated by relatively long pulses (several tens of ps) were measured using a VUV streak camera by Faldon {\it et al.} \cite{faldon} and Starczewski {\it et al.} \cite{starczewski}.  In order to measure the duration of harmonic pulses in the femtosecond regime, Schins {\it et al.} \cite{schins} developed  a cross-correlation method in which they ionize helium atoms by combining
 two pulses: the fundamental (800
nm, 150 fs) from a Ti:Sapphire laser, and its 21-st harmonic (38 nm).
These pulses generate characteristic electron spectra whose sidebands scale as the cross-correlation function which can be mapped out by varying the delay between the two pulses. Using variants  of this method  Bouhal {\it et al.}
\cite{bouhal}, and Glover {\it et al.} \cite{glover} were able to measure the duration
of 21-st to 
27-th harmonics within a sub-picosecond accuracy. For instance, in Ref.
\cite{bouhal} for a fundamental pulse of 190 fs FWHM, durations of 100$\pm$30 fs and $150\pm30$ were found for the FWHM of 21-st and 27-th harmonic, respectively. 

Concerning the spectral properties of the individual harmonics, the blue shift due to ionization has been reported in Ref. \cite{suedois}, whereas spectral 
properties of harmonics generated by chirped pulses have been discussed in Ref. \cite{kapteyn}.

\subsubsection{Theory}

The theoretical description of spatial distributions and temporal profiles of harmonics requires to combine a reliable  single atom theory that describes 
the nonlinear atomic response to the fundamental field, with a propagation code that accounts for phase matching, dispersion etc. Peatross {\it et al.}
have studied the spatial profiles of low-order harmonics in the loosely focused regime \cite{peatheor,peatheor2}. Muffet {\it et al.} \cite{muffet}
 modeled the results of Ref. \cite{tisch}, and showed that, depending on the focusing conditions, substructures could be either due to ionization or to resonances in the intensity dependence of the atomic phase. Temporal profiles of low-order harmonics were
discussed in Ref. \cite{faldon,starczewski}. Rae {\it et al.} \cite{rae} performed calculations outside the slowly varying envelope approximation by solving simultaneously the equations for the atomic dynamics and propagation, using a one dimensional approximation. Temporal and spectral profiles were studied in the strongly ionizing regime. 

In the series of papers \cite{PRL,opus}, we have developed a single atom theory  which is a quantum-mechanical version of the two-step model of Kulander {\it et al.} \cite{Kulsilap}
and Corkum \cite{Corkum}. This theory has been combined with
the theory of HG by macroscopic media \cite{propag} to describe experimental results in a realistic manner.  In a recent Letter \cite{prlphase}
we have stressed the 
role of the dynamically induced phase of the atomic polarization in phase matching and propagation processes. In particular we have demonstrated the possibility of controlling the spatial and temporal coherence of the harmonics by changing the focusing conditions of the fundamental.
We have performed numerical simulations of the angular distributions. The simulated profiles reproduced remarkably well the experimental trends and are thus used to interpret them in Ref. \cite{salier2}. The role of the intensity dependent phase of atomic dipoles was elaborated in more detail in Ref. \cite{phase} (see also \cite{kan}). In the Ref. \cite{phrev} we present a
short review of the various consequences of the intensity dependent phase.

\subsection{The scope of the present review}

As already stated above, the knowledge of the coherence properties of high harmonics is of major importance both for applications and from the fundamental point of view. Although experimental and theoretical work has been already
devoted to this subject, a systematic study of the spatial and temporal
coherence 
properties of harmonics is still lacking. In particular, the implications of the existence of a phase of the harmonic dipole have not been fully explored.
The aim of this review is to present a detailed theoretical study of this important subject. The theoretical approach used in this paper is very well established,
and has been confronted numerous times with experimental results with great success. Some  of the results presented in this review
have been published before and compared with experiments. Many of the results, however, are either new, or have been only reported in the PhD thesis of P. Sali\`eres \cite{these}.
Nevertheless, we feel that very soon these results will find their experimental
confirmation. 

To some extent this review has a character of a case study, i.e.
we discuss here in great detail quantitative coherence properties of
a specific high harmonic in specific conditions etc. It is important, however, to keep in mind that the presented 
results are not only qualitatively, but also quantitatively valid in more general cases. We hope  that these results will turn out usefull for anybody interested in applications of harmonics in general, and their extraordinary coherence properties in particular.

The plan of the review is the following. 
In Chapter 2, we give an overview of theoretical approaches and
describe our theoretical method: the single atom
theory based on strong field approximation,
and the propagation equations. Since our theory has been discussed in detail in other publications, we limit ourselves to present the final 
expressions that we use for calculations of the physical quantities.
In Chapter 3 we discuss the phase matching problem stressing the role of the 
dynamically induced phase of the atomic polarization \cite{prlphase,phase}.
The combined effects of this phase and the phase of the fundamental beam
depend on the atomic jet position relative to the focus. We investigate the influence of the jet position on the conversion efficiency. We show how the phase matching effects modify the cut-off law.

Chapter 4 starts with a short section devoted to the general definitions of the degree of coherence and characteristics of partially coherent beams.
 We then concentrate on the emission profiles and quality of the wavefronts of harmonics, both in the near-field and in the far-field zones.
We also present calculations of the degree of spatial coherence of the harmonics.

In Chapter 5 we turn to the discussion of the temporal and spectral coherence.
We show how the intensity dependence of the phase of the atomic polarization
leads to a temporal modulation of the harmonic phase and to a chirp of its frequency.  In both Chapters 4 and 5, we relate our theoretical findings to experimental results. In particular, we use the parameters corresponding to experiments of Ref. \cite{salier2}, and discuss systematically
the  dependence of the coherence properties of harmonics on the focus
position of the fundamental. This parameter, as shown in Ref. \cite{prlphase},
allows us to control the degree of coherence; optimal coherence properties are obtained when the fundamental is focused sufficiently before the 
atomic jet. In the final part of Chapter 5, we discuss the possibility of optimizing and/or controlling the temporal and spectral properties of harmonics by making use of the dynamically induced chirp: temporal compression with a grating pair and spectral compression by using
a chirped fundamental pulse. These ideas are confronted with the recent experiment of Zhou {\it et al.}
\cite{kapteyn}, and to the theoretical proposal of Kulander \cite{kenpriv}. 

In Chapter 6 we discuss future applications of harmonics with the special emphasis on their coherence properties. In particular we focus on 
applications in interferometry and on the short pulse effects. We 
discuss the possibility of generating and applying attosecond pulses. Finally, we conclude in Chapter 7.

\section{Theory of harmonic generation in macroscopic media}

The theory of harmonic generation in macroscopic media must necessarily contain two components: (i) a single atom theory that describes the response of an atom to the driving fundamental laser field, and (ii) a theory of propagation of the generated harmonics in the medium. In this Chapter we outline various
 approaches to describe these two components of the theory.

\subsection{Single atom theories}

The single atom theory should  describe the single atom response to a time-varying field of arbitrary intensity, polarization and phase.
In other words, it should allow to calculate
the induced atomic polarization, or dipole acceleration, which then can be 
inserted as a source in the propagation (Maxwell) equations.
  In principle it is sufficient to describe the atomic 
response in the framework of the single active electron (SAE) 
approximation  (cf. \cite{kul88}; for a discussion of two-electron effects see for instance \cite{saneb,erhard,taylor}). Also
for relatively long laser pulses (of duration down to $\simeq 50-100$ fs for Nd-Glass or Ti-Sapphire lasers) one can use the adiabatic approximation, i.e. calculate the atomic response for the field of constant intensity, and only at the end integrate the results over the ``slowly'' varying envelope of the laser pulse. The discussion of the validity of the adiabatic approximation is presented in more detail in Chapter 5.

There are essentially four methods that have been used to solve the problem
of the single atom response :

\begin{itemize}
\item{\it Numerical methods} These methods allow to solve the time dependent Sch\"odinger equation (TDSE) describing an atom in the laser field. Since 
(at least in the adiabatic case) the field oscillates periodically, one of the possible approaches is to use the Floquet analysis (\cite{potvliege}; for a recent review see \cite{joachain}),
but the direct integration of the TDSE is far more often used (for a review see \cite{kulrev,Protopapas}). In 1D such integration can be performed using either 
the finite element (Crank-Nicholson), or split operator techniques; in the context of harmonic generation it has been  first used  by the Rochester group \cite{eberly},
but then employed by many others as a test method.

In 3D the numerical 
 method has been initiated by Kulander \cite{kul87,kul88}, who used a 2D finite element (``grid'') method. Pretty soon it was realized that basis expansion methods that employ the symmetry of the problem (i.e. the spherical symmetry of a bare atom, or the cylindrical symmetry of an atom in the linearly polarized field) work much better \cite{devries,lagutta}. Modern codes use typically
expansions in angular momentum basis, and solve the coupled
set of equations for the radial wave  function using finite grid methods (cf. \cite{Krause}), Sturmian expansions \cite{antstur,antst2} or B-spline expansions \cite{cormier1,cormier2}. Most of those 
codes are quite powerful and allow to calculate the
atomic response directly without adiabatic approximation (cf. \cite{kenpriv}).
Unfortunately,  they are quite time consuming, and it is therefore
very hard to combine the results obtained from the numerical solutions of TDSE with the propagation codes. 
The reason is that the single atom response in the physically interesting regime is typically a rapidly varying function of the laser intensity and other laser parameters. The propagation codes thus require very detailed data from single atom codes. This problem becomes even more 
serious in the absence of cylindrical symmerty;  real 3D numerical codes
(such as the ones describing generation by elliptically polarized fields)
have been developed only recently
(cf. \cite{antstur,antst2,protoq}), and obviously are even  more time and memory
consuming. 

Nevertheless, many seminal results concerning harmonic generation has been obtained using direct numerical methods:
from the first observation of the $I_p+3U_p$ law \cite{Krause}, to the recent proposal of attosecond pulse generation \cite{kenpriv}. Particularly interesting are
the contributions of the Oxford-Imperial College group (for a review see \cite{Protopapas}) that concern among others
HG by short wavelength lasers \cite{preston,sanp}, pulse shape and blue shifting effects \cite{Watson}, role of strong ionization in HG \cite{rae2},
temporal aspects of harmonic emission \cite{rae,wavepack}, and the generation
from the coherent superposition of atomic states \cite{super,sanpml}. The TDSE method has also been applied to molecules aligned in the laser field \cite{Zuo,Krause2,Plummer}.

\item{\it Classical phase space avaraging method} A lot of useful information about high harmonic 
generation processes can be gained from a purely classical analysis of  the electron driven by the laser field. In order to mimic quantum dynamics, classical Newton equations are solved here for an ensemble of trajectories generated from an initial electron distribution in the phase space. This distribution is supposed to mimic the true quantum initial state of the system, so that averages over this distribution  are analogs of quantum averages. Such approach has been  developed in the context of HG by Maquet and his collaborators
\cite{banda,veniard} (see also \cite{balthese}).
 
\item{\it Strong field approximation} As already mentioned, the seminal paper
on the $I_p+3U_p$ law \cite{Krause} stimulated the formulation
of the ``simple man's theory'' \cite{Kulsilap,Corkum}. Originally, this
theory has been formulated as a mixture of quantum and classical
elements: first,  the tunneling  of the electron
out from the nucleus was described using the standard
ADK \cite{A,D,Kra} theory of tunneling ionization. The subsequent 
oscillations of the electron in the laser field were described using classical mechanics.  Finally, electron recombination back to the ground state
was calculated using the classical cross-section for the collision and the 
quantum mechanical recombination probability. 

A fully quantum mechanical theory, that recovers the ``simple man's theory'' in the semi-classical limit was formulated soon after \cite{PRL,opus}. This theory is
 based on the strong field approximation (SFA) to the TDSE. It is 
 a generalization of the  Keldysh-Faisal-Reiss approximation \cite{K,F,R}, 
applied to the problem of harmonic generation. It was for the first time formulated in the context of harmonics by Ehlotzky
\cite{ehlo}; it is also strongly related to the, so called, Becker model
of an atom with a zero range pseudo-potential
interacting with the laser field (see below). 

In our formulation, the theory is based on the  following assumptions: (i) it neglects all bound states of the electron in an atom with exception of  the ground state; (ii) all the states 
in the ionization continuum are taken into account, but in their dynamics only
the part of the Hamiltonian responsible for the oscillations of the free electron in the laser field is kept. Technically, we disregard all off-diagonal continuum-continuum transitions that change  electronic velocity. With this two assumptions, the TDSE becomes exactly soluble, and the resulting solutions are valid provided $U_p\ge I_p$. In most of applications we treat the electronic
states in the continuum as Volkov plane waves, which additionally limits 
the validity of our approximation to electronic states with high kinetic energy,
and thus to the description of the generation of high harmonics (with photon
energy $\ge I_p$). It is worth stressing, however, 
that a heuristic scheme of accounting for Coulomb potential effects in the continuum within the framework of SFA was proposed by \cite{ivabra}.

Originally our method has been formulated for linearly polarized laser fields in the adiabatic (slowly varying intensity) approximation
\cite{PRL,opus}. We have since then generalized it to elliptically polarized fields \cite{ellip}, two-color fields \cite{Mette}, and
 the fields with periodically time-dependent polarization \cite{Antoine2}, where all of those results were obtained in the adiabatic approximation. Finally, the method has been generalized  to the fields with arbitrary time-dependence
without adiabatic approximation (see Chapter 5). We have also performed intensive studies of the semiclassical approximation applied to our method in order to understand the harmonic emission in terms of semiclassical electron trajectories and Feynman path integrals \cite{phase,phrev}. These studies 
were essential for understanding the role of the intensity dependent phase of the nonlinear atomic polarization \cite{prlphase}. They allowed also 
to understand  the
mechanism of electron trajectory selection in propagation, responsible for the generation of attosecond pulse trains \cite{Antoine,Balcou}. 
The advantage of our method, apart from its very transparent physical sense,
 is that it gives
partially analytic results, allowing rapid calculation of the very precise data required for propagation codes. Last, but not least, our method combined with propagation codes gives results in very good agreement with experiment; it has become the standard theoretical method of analysis of experimental data
in the Saclay and Lund groups; it is also used by other groups \cite{kondo,doerr}.

\item{\it Pseudo-potential model} Many important results in the theory of harmonic generation have been obtained by Becker and his collaborators who have 
solved exactly (and to a great extent analytically) the zero range
pseudo-potential model \cite{bec1}. In this model the electron is bound to the nucleus via the potential
\begin{equation}
V(\vec r)=\frac{2\pi}{\kappa m }\delta(\vec r)\frac{\partial}{\partial r}r,
\end{equation}
where $m$ is the electron mass.
This potential supports a single bound state with the energy
$-I_p=-\kappa^2/2m$. 

This model, originally formulated in the case of a linearly polarized field, was also extended to one-color \cite{becker}, and two-color
\cite{Long} fields with arbitrary polarization. It was also used to study the polarization properties of harmonics generated by elliptically polarized fields \cite{lohr}. As our SFA theory
\cite{opus,ellip}, Becker's model 
 may rigorously account for the ground state depletion
\cite{bec2}.  Structures in the harmonic spectra were associated in this model to the above-threshold  ionization channel closings \cite{bec3}, rather than with quantum interferences between the contributions of different  electronic trajectories \cite{phase}. Nevertheless, Becker's model leads practically
to  the same final formulas for the induced atomic dipole moment as our SFA
theory, and to very similar results (the small discrepancies are
caused by additional approximations  used for numerical 
elaboration of final expressions; for detailed comparison of the two
models see \cite{compbeck}). Becker's model has also been used by several groups
to analyze experimental data \cite{harg,Hannover,paulus}.
\end{itemize}

\subsection{Single atom response in the strong field approximation}

In this section we present explicit formulas describing the 
response of a single atom to the laser pulse in the
strong field approximation. Since the details of our version
of the SFA can be found in the series of Refs. 
\cite{PRL,opus,prlphase,phase,ellip}, we limit ourselves to  present the relevant expressions and to discuss their physical sense.

Within our approach we obtain an approximate solution of the time dependent Schr\"odinger equation that describes an atom in the strong electric
field of a laser of frequency $\omega$ in the single active electron approximation. The knowledge of the time dependent wave function $|\Psi(t)\rangle$
allows us to calculate the time-dependent dipole moment 
 ${\vec x}(t)=\langle
\Psi(t)|{\vec x}|\Psi(t)\rangle$ in the form of a generalized Landau-Dyhne
formula \cite{Landau,Landau2} (we use atomic units)
\begin{eqnarray}
	{\vec x}(t)&=& i\int_0^{\infty}d\tau\;\left(\frac{\pi}
{\nu + i\tau/2}\right)^{3/2}\;
	 {\vec  d}^*({\vec p}_s-{\vec A}(t))\;
\exp{\left(-iS({\vec p}_s,t,\tau)\right)} \;\nonumber \\
\times&&{\vec{\cal E}}(t-\tau){\cdot \vec d}({\vec p}_s-{\vec A}(t-\tau))\exp(-\Gamma t)
	+ c.c.,
\label{xfinsp1}
\end{eqnarray}
where $\nu$ is a positive regularization constant, $\vec A(t)$ denotes
the vector potential of the electromagnetic field, $\vec{\cal E}(t)=
-\partial \vec A(t)\partial t=(E_{1x}\cos(\omega t), E_{1y}\sin(\omega t),0)$ is the electric field (polarized elliptically, in general), whereas 
 $S({\vec p},t,\tau)$ is the {\it quasi--classical action}, describing the motion of an electron moving in the
laser field with a constant canonical momentum $\vec p$,
\begin{equation}
S({\vec p},t,\tau)= \int_{t-\tau}^t dt''\,
	\left(\frac{({\vec p}-{\vec A}(t''))^2}{2}+I_p\right),
\label{action}
\end{equation}
with $I_p$ denoting the ionization potential.  In expression (\ref{xfinsp1})
we have already performed (using the saddle point method)
the integral over all possible values of the
momentum $\vec p$ with which the electron is born in the continuum. For this reason the integral in Eq.  (\ref{xfinsp1}) extends only over the possible
{\it return} times of the electron, i.e. the times it spends in the continuum
between the moments of tunneling from the ground state
to the continuum and recombination back to the ground state. The saddle point
value of the momentum (that is at the same time
 a stationary point of the quasi-classical action)
is 
\begin{equation}
{\vec p}_s={\vec p}_s(t,\tau)=
\int_{t-\tau}^t dt''\;{\vec A}(t'')\tau.
\label{pstat}
\end{equation}
Note the characteristic prefactor $(\nu +i\tau/2)^{-3/2}$ in  (\ref{xfinsp1}) coming from the effect of quantum diffusion. It cuts off very efficiently the contributions from large $\tau$'s and allows us to extend
the integration range from $0$ to infinity.

 The field--free dipole transition element from the ground state to the continuum state characterized by the momentum $\vec p$  can
be approximated  by \cite{opus,bethe}
\begin{equation}
{\vec d}({\vec p}) = i\frac{2^{7/2}\alpha^{5/4}}{\pi}
\frac{{\vec p}}{({\vec p}^{\ 2}+\alpha)^3},
\label{dhydr}
\end{equation}
with $\alpha=2I_p$, for the case of hydrogen--like atoms and transitions from $s$--states.

Finally, $\Gamma$ is the ionization rate from the ground state. In the framework of our theory it can be represented as twice the real part of the time-averaged
complex decay rate
\begin{eqnarray}
  \gamma(t) &=& \int_0^{\infty}d\tau \;\left(\frac{\pi}
{\nu + i\tau/2}\right)^{3/2}\;
\vec{\cal E}^*(t)\cdot
\vec d^*({\vec p}_s-{\vec A}(t))\;
\exp(-iS({\vec p}_s,t,\tau))\; \nonumber \\
&\times &
{\vec{\cal E}}(t-\tau)\cdot
{\vec d}({\vec p}_s-{\vec A}(t-\tau)).
\label{gammasp}
\end{eqnarray}

Note that both expressions (\ref{xfinsp1}) and (\ref{gammasp}) have the characteristic form of semiclassical expressions that can be analyzed in the spirit of Feynman path integral: they contain (from right to left)
transition elements from the ground state to the continuum at $t-\tau$, propagator in the continuum
proportional to the exponential of $i$ times the quasi-classical  action, and the final transition elements from the continuum to the ground state. Applying the saddle point technique to calculate the integral over $\tau$ (and $t$ if one calculates the corresponding Fourier components or time averages), one can transform both expressions into the sums of contributions corresponding to quasi-classical electron trajectories, characterized by the moment when the electron is born in the continuum $t_s-\tau_s$, its canonical momentum $\vec p_s(t_s,\tau_s)$ (see Eq. (\ref{pstat})), and the moment when it recombines $t_s$ \cite{opus,phase}.
Note, however, that due to the fact that we deal here with the tunneling
process (i.e. passing through the classically forbidden region), these trajectories will in general be complex. Typically, only the trajectories with the shortest return times ${\rm Re}(\tau)$ contribute significantly to the expression (\ref{xfinsp1}); there are two such relevant trajectories with return times shorter than one period, i.e.  $0<{\rm Re}(\tau_1)<{\rm Re}(\tau_2)<2\pi/\omega$.

Note also that the dipole moment (\ref{xfinsp1})
can be written in the form
\begin{equation}
{\vec x}(t) = \sum_{q\ {\rm odd}}{\vec x}_q e^{-iq\omega t -\Gamma t} +{\rm c.c.}
\label{repre}
\end{equation}
where $x_q$ denote Fourier components. They can be calculated either directly from Eq. (\ref{xfinsp1}) using a  Fast Fourier Transform, or analytically
as discussed in Ref. \cite{ellip}.

It is important to remember that both expressions (\ref{xfinsp1}) and (\ref{gammasp}) result from the single active electron approximation. Before inserting these expressions into the propagation equations, one has to account for the contributions of all active electrons, and replace Eqs. (\ref{xfinsp1}) and (\ref{gammasp}) by the total dipole moment and the total ionization rate
that are given by the sums of the corresponding (independent) contributions of 
all active electrons. In the case of Helium (two $s$ electrons in the ground state), this amounts to multiplying both expressions by the factor two. In the case of other noble gases (six $p$ electrons in the ground state, two in each of the $m=-1,0,1$ states) the procedure is more complex. Both expressions should be replaced by two times the sums of contributions of the given magnetic quantum number $m=-1,0,1$; each of those contributions should be calculated replacing Eq. (\ref{dhydr}) by 
 an appropriate field--free dipole matrix element describing the transition from the $l=1,m=-1,0,1$ states to the continuum. 

Fortunately, the dependence of the dipole moment and ionization rate on the details of the ground state wave function is rather weak, and typically reduces to an overall prefactor \cite{opus,ellip}, that determines the strength of the dipole, but not the form of its intensity dependence. For these reasons,
in most of the calculations for noble gases other than Helium, we still use the $s$-wave function to describe the ground state (\ref{dhydr}), but multiply the results by an effective number of active electrons, $n_{el}$, that is equal to that ranges between $2<n_{el}\simeq 4< 6$ for other noble gas atoms. Total ionization rates of Helium and Neon calculated with $n_{el}=2, \simeq 4$, respectively,  agree very
well with the ADK ionization rates \cite{A,D,Kra}.

\subsection{Propagation theory}

In order to calculate the macroscopic response of the system, one has to solve 
the Maxwell equations for the fundamental and harmonic fields. This can be done using the slowly varying envelope and paraxial approximations. The fundamentals of such an approach have been formulated by \cite{propag}. Several
groups have used similar approaches to study the effects of phase matching, and to perform direct comparison of the theory with experiments \cite{muffet,peatheor,peatheor2,rae1,rae}. To our knowledge, the most systematic studies of these sort have been so far realized by the Saclay--Livermore--Lund collaboration.

In a series of papers we have studied propagation and phase matching effects 
in the context of the following problems:
(i) phase matching enhancement in nonpertubative regime \cite{anneliv}; (ii) phase matching effects in tight focusing conditions \cite{annelom,balc93}; 
(iii) shift of the observed cut-off position \cite{PRL}; (iv) density dependence of the harmonic generation efficiency \cite{altucci}; (v) harmonic generation by elliptically polarized fields \cite{ellip}; (vi) harmonic generation by two-colored fields \cite{Mette}; (vii) coherence control of harmonics by adapting the focusing conditions \cite{prlphase}; (viii) influence of the experimental parameters on the harmonic emission profiles \cite{salier2}; (ix) generation of attosecond pulse trains \cite{Antoine}; (x) generation of attosecond pulses by laser fields with time-dependent polarization \cite{Antoine2}, (xi) interference  of two overlapping harmonic beams \cite{wahl,zerne}, and others. We have also formulated generalized phase matching conditions that take into account the intensity dependent phase of the induced atomic dipoles \cite{Balcou}.

\subsection{Macroscopic response}

In this section we present the Maxwell equations for the fundamental and harmonic fields used in our above mentioned studies. Using the slowly varying envelope and paraxial approximations, the propagation equations can be reduced to the form (here we use SI units)
\begin{eqnarray}
& &\nabla_\bot ^2 \vec E_1(\vec r,t)+2ik_1^0{\partial \vec E_1(\vec r,t) \over\partial z} + 2k_1^0 \delta k_1(\vec r,t) \vec E_1(\vec r,t)= 0 \label{prop6} \\
& &\nabla_\bot ^2 \vec E_q(\vec r ,t)+2i k_q^0{\partial \vec E_q(\vec r,t) \over\partial z}+  2 k_q^0 [\Delta k_q^0(z)+ \delta k_q(\vec r,t)]\vec E_q(\vec r,t)\nonumber \\
& &= - {q^2\omega^2\over \epsilon_0 c^2} \vec P_q^{NL}(\vec r,t) ,\label{prop7}
\end{eqnarray}
where $\vec E_1(\vec r,t)$, and $\vec E_q(\vec r ,t)$ denote the slowly varying 
(complex) envelopes of the fundamental and harmonic fields respectively, $k_q^0=
q\omega/c$, whereas the rest of the symbols is explained below. The slow time dependence in the  above equations accounts for the temporal profile of the fundamental 
field that enters Eq. (\ref{prop6}) through the boundary condition
for $\vec E_1$. The solutions of the propagation equations for given $t$ have therefore
to be integrated over $t$.

The terms containing $ \Delta k_q^0(\vec r,t)$ describe dispersion effects due to the linear polarisability of atoms, and can be in fact neglected in the
regime of parameters considered (low density). The terms proportional to $ \delta k_q(\vec r,t)= -e^2{\cal N}_e(\vec r,t) /2mqc\omega$, with $e$ denoting the  electron
charge, $m$- its   mass, and ${\cal N}_e(\vec r,t)$ the electronic density,
describe the corrections to the index of refraction due to ionization; here the ionic part of those corrections is neglected. The electronic density
is equal to the number of ionized atoms, i.e. 
\begin{equation}
{\cal N}_e(\vec r,t)= {\cal N}_a(z)\left[1-\exp\left(-\int_{-\infty}^t
\Gamma(|\vec E_1(\vec r,t')|)\,dt'\right)\right]
\end{equation}
where ${\cal N}_a(z)$ is the initial density of the atomic jet, and $\Gamma(|\vec E_1(\vec r,t')|)$ is the total ionization rate, which 
takes into account the contributions of all active electrons calculated from Eq. (\ref{gammasp}) using $\Gamma=2 {\rm Re}[\int_0^T\gamma(t)dt/T]
$ with $T=2\pi/\omega$ for an instantaneous and local value of the electric field envelope  $\vec E_1(\vec r,t')$. Note that since $\Gamma$ depends functionally on $\vec E_1(\vec r,t')$, Eq. (\ref{prop6}) is a nonlinear integro-differential equation; it has to be solved first, and its solution is used then to solve Eq. (\ref{prop7}).

Finally, the Fourier components of the atomic polarization are given
by
\begin{equation}
\vec P_q^{NL}(\vec r,t)=2{\cal N}_a(z) \vec x_q(\vec r,t)e^{iq\phi_1(\vec r, t) }\exp\left(-\int_{-\infty}^t
\Gamma(|\vec E_1(\vec r,t')|)\,dt'\right),
\label{polar}
\end{equation}
where 
 $\vec x_q(\vec r,t)$ denote the  harmonic components of the total atomic dipole moment, that includes the contibutions of all active electrons 
 calculated from Eq. (\ref{xfinsp1}) for a field 
($|E_{1x}| \cos(\omega t), |E_{1y}|\sin(\omega t), 0$).
 The factor of 2 arises from different conventions used in the definitions of $\vec P_q^{NL}$ and $\vec x_q$. 
Finally, $\phi_1(\vec r,t)$ represents the phase of the laser field envelope $\vec E_1 (\vec r,t)$, obtained by solving the propagation equation for the fundamental.

\section{Phase matching}

In this chapter, we study how the harmonic field builds up in the medium, and is influenced by the dynamically induced phase of the atomic polarization. 
In all the calculations presented in this work, we try to mimic the experimental conditions of Refs. \cite{salier2,prlphase}. The 825 nm wavelength laser is assumed to be Gaussian in space and time, with a 150 fs full width at half maximum (FWHM). 
The laser confocal parameter $b$ is equal to 5 mm and the focus position is located at $z=0$. The generating gas is neon and the atomic density profile is a Lorentzian function centered at $z$ with a 0.8 mm FWHM, truncated at $z$ $\pm 0.8$ mm. 

\subsection{Source of the harmonic emission}

The generic intensity dependence of the dipole strength (i.e. absolute value squared) and phase for the 45th harmonic generated by a single neon atom is shown in Fig. \ref{dipole}. At low intensity, when the harmonic is in the cutoff region, the dipole strength (dashed line) increases rather steeply with laser intensity and the phase (solid line) decreases linearly. 
When the harmonic enters the plateau region, the strength saturates and exhibits many interferences, while the phase decreases twice as fast as in the cutoff region, predominantly linearly but with superimposed oscillations.
The regular behaviour of the dipole in the cutoff region is due to the existence of only one main electron trajectory leading to the emission of the considered harmonic. On the contrary, in the plateau, there are at least two relevant trajectories whose contributions interfere, resulting in a perturbed dipole.
In this region the dominant contribution is typically
from the trajectory with the longer return time $\tau_2$ (see section 2.2) -- it is the action along this trajectory  
which determines the mean slope of the intensity dependence of the phase.

One should mention, however, that the mean slope of the intensity dependence 
is  not always a good measure of the phase variations in the macroscopic medium.
The reason is that propagation and phase matching may lead to a single trajectory selection that depends on the relative position  of the atomic jet with respect to  the focus \cite{Antoine}. For the case when the jet is before, or close to the focus, phase matching usually selects the dominant 
trajectory with the return time $\tau_2$, and the intensity dependence of the phase after  the selection follows the mean slope of the total phase. When the jet is after the focus, however, and when intensity in the jet is high enough,
the trajectory with the return time $\tau_1$ is selected, and the slope of the intensity dependence is much weaker than the mean slope of the total phase.
The latter case will not be considered in the following; we shall discuss only the situations when the mean slope of the phase can be 
used to characterize the intensity dependence of the phase in the macroscopic medium.

This intensity dependence of the dipole is quite universal for sufficiently high order harmonics, and depends weakly on harmonic number, laser wavelength, generating gas. These parameters only determine the position of the plateau-cutoff transition. In the following, we will thus concentrate on the study of the 45th harmonic generated in neon, keeping in mind that its behaviour is very general.

\subsection{Dynamically induced phase of the atomic polarization} 

Harmonic generation is optimized when phase-matching is achieved, i.e. when the difference of phase between the generated field and the driving polarization $\Delta \Phi = \Phi_q - \Phi_{pol}$ is minimized over the medium length, allowing an efficient energy transfer. In the case of plane waves and low order harmonics, this results in the well known phase matching condition on the wave vectors: $\Delta k = k_q-qk_1 \approx 0$. However, for the very high orders discussed here, the variation of the phase of the polarization is much more important than that of the induced field, so that matching phases amounts to minimizing the phase variation of the driving polarization.
 One of the main causes of variation of the polarization in the medium is due to the rapid variation of the dipole phase with intensity. It induces a spatially and temporally dependent phase term in the medium, which influences the generation of the macroscopic field.

\begin{figure}[h]
\vspace*{0.8cm}
\centerline{
\psfig{angle=90,height=6cm,file=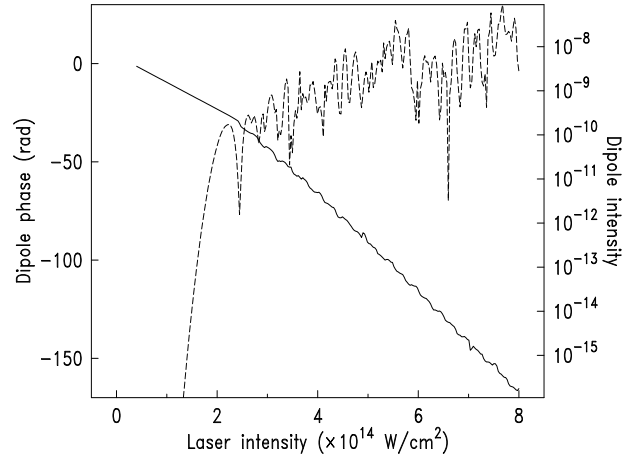}}
\vspace*{0.8cm}
\caption{45th single neon atom harmonic intensity (dashed line) and phase (solid line) as a function of the laser intensity.
\label{dipole}}
\end{figure}

\begin{figure}[h]
\vspace*{0.8cm}
\centerline{
\psfig{angle=90,height=6cm,file=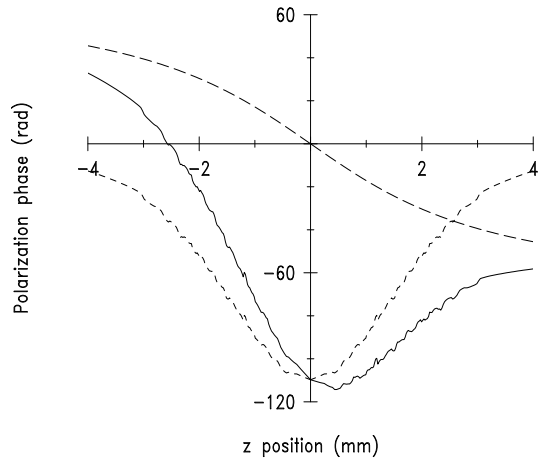}}
\vspace*{0.8cm}
\caption{Phase of the polarization on the propagation axis (solid line). 
The long-dashed line indicates the term due to the propagation of the fundamental, and the short-dashed line the dipole phase for a 
peak intensity of 6 $\times 10^{14}$ W/cm$^2$.
The laser propagates from the left to the right.
\label{phaspol}}
\end{figure}

 Let us consider this variation on the propagation axis at the maximum of the pulse temporal envelope for the 45th harmonic. 
It is induced by the variation of the intensity $I(z)= I_0/(1+4z^2/b^2)$, and is shown by the short-dashed line in Fig. \ref{phaspol} for a peak intensity, i.e. at best focus and at the maximum of the pulse envelope, of 6 $\times 10^{14}$ W/cm$^2$. 
The variation of the dipole phase is less rapid outside the interval $[-3$ mm, $+3$ mm~$]$, when the intensity on axis corresponds to the cutoff region ($I(z) \le  2.4 \times 10^{14}$ W/cm$^2$, see Fig. \ref{dipole}). 
The other important contribution to the polarization phase is a propagation term induced by the phase shift of the Gaussian fundamental field in the focus region, equal to $-q\; \tan^{-1}(2z/b)$, $q$ denoting the 
process order. This function is shown in long--dashed line in Fig. \ref{phaspol}. There are other possible causes of variation of the polarization phase, such as atomic or electronic dispersion, but they are negligible in the conditions considered here.
The total phase of the nonlinear polarization is represented in Fig. \ref{phaspol} by the solid line. In the region $z < 0$, the variations of both phases add, leading to a rapid decrease of the total phase. In the region $z >0$, they have opposite signs, and almost 
compensate when the intensity on axis corresponds to the cutoff region. 
Consequently, phase matching strongly depends on the position of the medium 
relative to the laser focus.
The best phase matching conditions {\it on axis} are those for which the phase 
variation of the polarization over the medium length ($\sim$ 1 mm) is minimal, 
i.e. when the laser is focused approximately 3 mm {\it before} the generating 
medium. Note that, at the minimum of the curve close to the focus, the superimposed oscillations are detrimental to a good phase matching. 

\begin{figure}[h]
\vspace*{0.8cm}
\centerline{
\psfig{angle=90,height=6cm,file=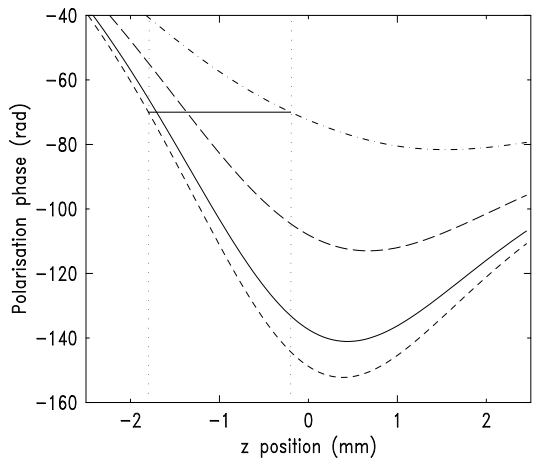}}
\vspace*{0.8cm}
\caption{Phase of the polarization for different radial positions relative to the propagation axis, for a peak intensity of 6 $\times 10^{14}$ W/cm$^2$: $r=0$ (short-dashed), $r=5\mu m$ (solid), $r=10\mu m$ (long-dashed) and $r=15\mu m$ (dot-dashed). 
The dotted lines indicate the edges of a gas jet placed in $z=-1mm$, and the horizontal solid line, a trajectory $r(z)$ that keeps the phase constant.
\label{phaspoloff}}
\end{figure}

So far, we have only considered phase matching on axis, which corresponds to centered harmonic profiles. However, good phase matching {\it off axis} can be realized in certain conditions. 
This is illustrated in Fig. \ref{phaspoloff} with the longitudinal variation of the polarization phase for different radial positions, from $r$ =0 to 15 $\mu$m (relative to the propagation axis) for a peak intensity of 6 $\times 10^{14}$ W/cm$^2$. Here, for clarity, the curves have been smoothed so that the superimposed oscillations in the plateau region do not appear. 
For a gas jet centered in $z=-1$ mm, it is possible to minimize the longitudinal phase variation by moving from one curve to the other, i.e. by going off the propagation axis. 
Along these favored directions, the quick variation of the laser intensity on axis (cause of the rapid decrease of the polarization phase) is avoided by going off axis. In these conditions, the harmonic field can build up efficiently in the plateau region. Note that a method allowing the systematic study of these generalized phase-matching conditions has been proposed in Ref. \cite{Balcou}. 

\begin{figure}[h]
\vspace*{0.8cm}
\centerline{
\psfig{angle=90,height=6cm,file=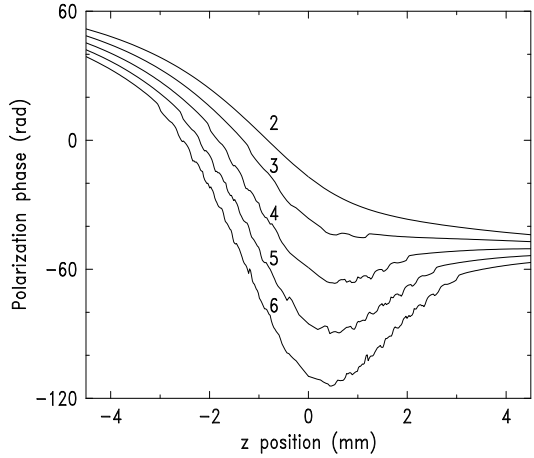}}
\vspace*{0.8cm}
\caption{Phase of the polarization on the propagation axis. From the top to the bottom, the laser intensity $I = $ 2, 3, 4, 5, 6 $\times 10^{14}$ W/cm$^2$.
\label{phaspolint}} 
\end{figure}

If we now consider another peak intensity, the shape of the total phase is modified. This is illustrated on axis in Fig. \ref{phaspolint} for several peak intensities, from 2 to 6 $\times 10^{14}$ W/cm$^2$.
 As the intensity increases, the induced phase becomes more and more important in determining the total phase variation near the focus, which departs more and more from the arctangent term. 

The optimal phase matching position on axis is observed at different $z$ depending on the peak intensity, since it always corresponds to the plateau-cutoff transition of the dipole (2.4 $\times 10^{14}$ W/cm$^2$).
Thus, for a given geometry, there will not be a {\it static} phase matching during the laser temporal envelope, but a continuous distortion of the build up pattern in the medium. This {\it dynamic} phase matching complicates the interpretation of the processes.

\subsection{Influence of the jet position on the conversion efficiency}

\begin{figure}[h]
\vspace*{0.8cm}
\centerline{
\psfig{angle=90,height=6cm,file=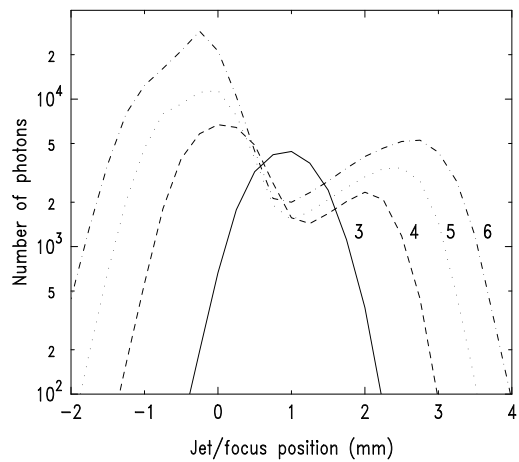}}
\vspace*{0.8cm}
\caption{Conversion efficiency for the 45th harmonic as a function of the position of the center of the jet relative to the focus, for peak intensities ranging from 3 to 6 $\times 10^{14}$ W/cm$^2$. The laser temporal envelope is square with a 150 fs width.
\label{convstat}}
\end{figure}

\begin{figure}[h]
\vspace*{0.8cm}
\centerline{
\psfig{angle=90,height=6cm,file=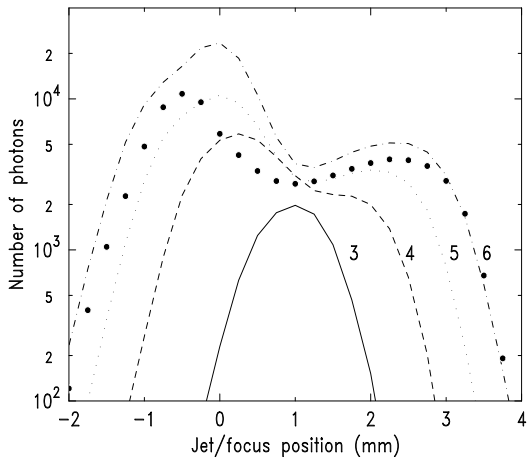}}
\vspace*{0.8cm}
\caption{Conversion efficiency for the 45th harmonic as a function of the position of the center of the jet relative to the focus, for peak intensities ranging from 3 to 6 $\times 10^{14}$ W/cm$^2$. The dots show the influence of ionization at 6 $\times 10^{14}$ W/cm$^2$. The laser temporal envelope is Gaussian with a 150 fs FWHM.
\label{convdyn}}
\end{figure}

Using the numerical methods described in Chapter 2, we perform the propagation of the generated harmonic fields in the medium, considering for the moment square laser temporal envelopes, i.e. static phase matching. 
In Fig. \ref{convstat}, we study the variation of the conversion efficiency for the 45th harmonic generation as a function of the position $z$ of the center of the atomic medium (relative to the laser focus placed in $z=0$), for peak intensities ranging from 3 to 6 $\times 10^{14}$ W/cm$^2$. The peak atomic density is 15 Torr. 

At low intensity, the curve presents only one maximum located in $z=1$ mm. Increasing the intensity, this maximum splits into two lobes that become more and more separated. This is the confirmation of the two optimal phase matching positions described above. The positions of the maxima $z>0$ correspond precisely in Fig. \ref{phaspolint} to the best phase matching positions {\it on axis}, i.e. to the plateau-cutoff transition of the dipole (2.4 $\times 10^{14}$ W/cm$^2$). The maxima occuring for negative $z$ correspond to optimized phase matching positions {\it off axis}, as shown in Fig. \ref{phaspoloff}. 
Note that all the curves are asymmetric compared to the focus position ($z=0$), and that the conversion efficiency in $z=1$ mm is larger at the lowest intensity.  However, this effect disappears when we consider harmonic generation by a Gaussian laser pulse, as shown in Fig. \ref{convdyn} for the same peak intensities as before. During the major part of the laser pulse, the intensity is in the cutoff region where the polarization amplitude drops, resulting in a lower conversion efficiency than for a square pulse. Except for this modification, the general behavior is the same as for square pulses, with enlarged peaks but similar positions and efficiencies. 

Note that if we take into account the ionization of the medium at 6 $\times 10^{14}$ W/cm$^2$ (dots in Fig. \ref{convdyn}), we find a marginal influence on the number of photons except very close to the focus position. In the following, we will thus neglect ionization in a first step, and include it afterwards when its effects are not negligible. 
Wahlstr\"om {\it et al.} \cite{suedois} have measured about 10$^5$ generated photons for the 45th harmonic in neon at 6 $\times 10^{14}$ W/cm$^2$. The difference of one order of magnitude with the results of
our simulations can be explained, at least partly, by the more optimized conditions used in their experiment (higher pressure and longer confocal parameter).

\begin{figure}[h]
\vspace*{0.8cm}
\centerline{
\psfig{angle=90,height=6cm,file=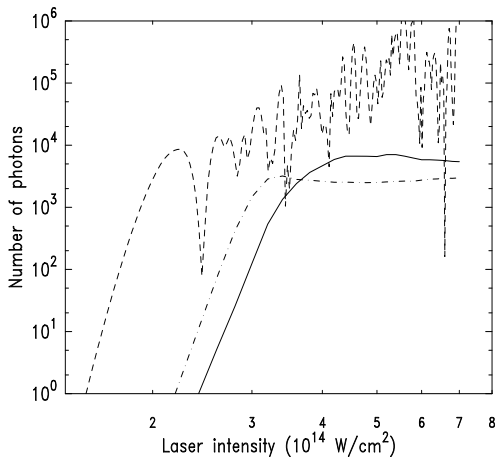}}
\vspace*{0.8cm}
\caption{Intensity dependence of the conversion efficiency for the 45th harmonic for a jet position in $z=0$ (solid line) and $z=1$ mm (dot-dashed line). Ionization is taken into account. The short-dashed line indicates the strength of the dipole moment (arb. units).
\label{intdep}}
\end{figure}

\subsection{Modified cut-off law}

The influence of the position of the focus relative to the gas jet on the
harmonic conversion efficiency has an interesting consequence on
the dependence of the harmonic yield as a function of intensity, for the different geometries. Fig. \ref{intdep} presents the intensity dependence of the conversion efficiency for the 45th harmonic for a focusing at the center of the jet (solid line, $z$=0). The comparison with the strength of the dipole moment (short-dashed line) indicates two main consequences of the propagation: the rapid variations in the plateau region are smoothed out, and the change of slope indicating the plateau-cutoff transition is shifted to a higher intensity. This shift is the same for higher order harmonics, and thus implies that propagation decreases the extent of the plateau of the harmonic spectrum compared to the single atom response, from a photon energy of $I_p+3.2U_p$ to about $I_p+2U_p$. This was observed experimentally by L'Huillier {\it et al.} \cite{PRL}, and explained in terms of the variation of phase matching with intensity.

If the laser was not focused in the jet, one would expect that the lower intensity experienced by the nonlinear medium would result in an even larger shift of the plateau-cutoff transition (recall that we are plotting the curves as a function of the peak intensity, i.e. {\it at the focus} and at the maximum of the pulse envelope). In Fig. \ref{intdep}, the intensity dependence for the case when the jet is located in $z=1$ mm is shown in dot-dashed line. Amazingly, the shift of the plateau-cutoff transition to higher intensities is less important than for $z=0$, corresponding to a cutoff law of about $I_p+2.3U_p$. This is a direct consequence of the optimization of phase matching at low intensity for this particular position, as shown in Fig. \ref{convdyn}. This effect is independent of the nonlinear order considered, and would indicate that the maximal extent of the plateau is not obtained for a focusing right into the jet, but rather slightly before it.

\section{Spatial coherence}

\subsection{Definition}
 
We recall here some notions of the theory of partial coherence (see, for example, Born and Wolf \cite{bornwolf}). The coherence of a beam is related to the correlation of the temporal fluctuations of the electromagnetic fields inside this beam. It is thus characterized by its mutual coherence function, defined for any two points inside the beam by: $\Gamma_{12} (\tau)=\left\langle E_1(t+\tau)E_2^*(t) \right\rangle $, where $E_1$, $E_2$ are the complex amplitudes of the electric field in these two points, and the angular brackets denote an appropriate time average (here, over the harmonic pulse).
The normalized form of the mutual coherence function is the complex degree of coherence:
\begin{equation}
\gamma_{12} (\tau)={\Gamma_{12}(\tau)\over \sqrt {\Gamma_{11}(0) \Gamma_{22}(0)}}={\left\langle E_1(t+\tau)E_2^*(t) \right\rangle \over \sqrt 
{\langle \vert E_1 \vert ^2\rangle 
\langle \vert E_2 \vert ^2\rangle}}, 
\label{cohdeg}
\end{equation}
whose modulus is known as the degree of coherence. The temporal coherence is described by $\gamma_{11} (\tau)$, while the spatial coherence is described by $\gamma_{12} (0)$. Note that the latter, despite its name, is related to the correlation {\it in time} of the fields emitted in two points. These quantities determine the ability of the fields to interfere, and can be measured in interferometry experiments: In Young's two-slit experiment, with both slits uniformly illuminated, $|\gamma_{12}(0)|$ at the slit positions is simply given by the fringe visibility, defined as $V=(I_{max}-I_{min})/(I_{max}+I_{min})$ (where $I_{max}$ and $I_{min}$ are the maximum and minimum intensities of the fringe pattern). 
The spatial coherence length of a beam at a given distance from its focus is defined as the length over which the degree of spatial coherence is larger than some prescribed value (between 0.5 and 0.9, depending on the authors, and on the coherence of their own source...).

Another important aspect related to the coherence of a beam is the quality of its wavefront, an aspect that is often confused with the preceding description. A beam is said to be ``diffraction limited" if the product of its spot size (at focus) and of its far-field spread (divergence) is of the order of the wavelength. This is realized when both the focal spot presents a reasonably regular amplitude variation, and the phase front at focus is very well behaved (typically plane). In particular, any distortion of the phasefront will result in a larger (e.g. N times) angular spread, and the beam will be called ``N times diffraction limited" \cite{siegman}.

In the following, we shall concentrate on the two focusing positions corresponding to well defined phase matching conditions at 6 $\times 10^{14}$ W/cm$^2$, namely $z=3$ mm (on axis) and $z=-1$ mm (off axis). For these extremal positions, on either side of the conversion efficiency curve (see Fig. \ref{convstat}), phase matching is mostly efficient close to the maximum of the laser temporal envelope, thus simplifying the study. Note that the main dependences of the harmonic emission profiles (laser intensity, nonlinear order, jet/focus position) have been intensively studied in Ref. \cite{salier2}, and compared successfully with experimental data. We here focus on the coherence properties.

\subsection{Study of the spatial coherence: atomic jet after the focus}

\begin{figure}[h]
\vspace*{0.8cm}
\centerline{
\psfig{angle=90,height=6cm,file=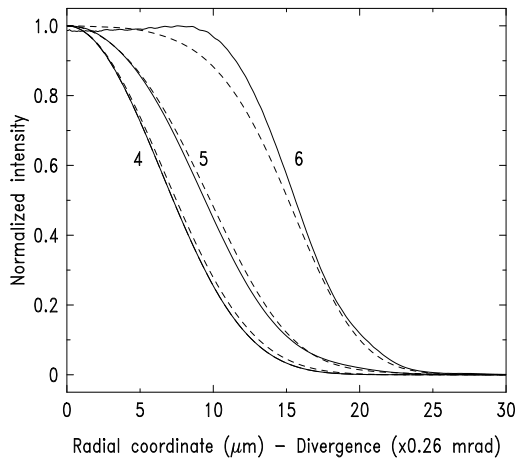}}
\vspace*{0.8cm}
\caption{Normalized spatial profiles for the 45th harmonic at the maximum of the pulse for a jet position in $z=3$mm, for intensities from 4 to 6 $\times 10^{14}$ W/cm$^2$. In solid lines are shown the profiles at the exit of the medium as a function of the radial coordinate, and in dashed lines, the far-field profiles (divergence).
\label{nfprof3}}
\end{figure}

First we study the characteristics of the harmonic beam in the near-field, i.e. at the exit of the medium in $z=3.8$ mm (the half width of the jet is 0.8 mm)
and at the maximum of the laser pulse. Fig. \ref{nfprof3} presents in solid lines the harmonic profiles corresponding to different intensities, from 4 to 6 $\times 10^{14}$ W/cm$^2$ (square pulses). The first two profiles are Gaussian, with 12 and 14 $\mu$m radius in 1/e$^2$ respectively, while the third is super Gaussian with a 20 $\mu$m radius. They are narrower than the fundamental (46 $\mu$m), but larger than the 7 $\mu$m  predicted by lowest order perturbation theory. These regular profiles result from a good phase matching on axis (see Chapter III) together with a regular intensity dependence of the amplitude of the dipole in the plateau-cutoff transition region. Increasing the intensity, this region is moved to the high density zone at the center of the jet, leading to a broadening and distortion of the profiles. 

\begin{figure}[h]
\vspace*{0.8cm}
\centerline{
\psfig{angle=90,height=6cm,file=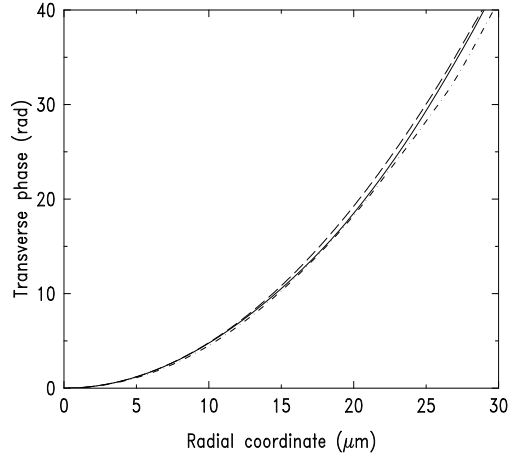}}
\vspace*{0.8cm}
\caption{Radial variation of the phases of the harmonic fields corresponding to the near-field profiles shown in Fig. \ref{nfprof3}: 4 $\times 10^{14}$ (dashed), 5 $\times 10^{14}$ (solid) and 6 $\times 10^{14}$ W/cm$^2$ (dot-dashed).
\label{nfphase3}}
\end{figure}

In Fig. \ref{nfphase3} we present the radial phases corresponding to these profiles. They all present a regular parabolic behavior, whose dependence is between $0.045r^2$ and $0.048r^2$ rad, $r$ being the radial coordinate in $\mu$m. 
To understand the origin of these curved phase fronts, let us consider the phase of the polarization at the exit of the medium. Given the low density, the harmonic field is  obviously not mainly generated there, but this gives an estimate of what happens in the medium and can be directly compared to the phase of the generated harmonic field. 
There are two main contributions to the polarization phase: the first one is the Gaussian fundamental field phase multiplied by the order: 
\begin{equation}
-q \arctan \left( 2z \over b \right)+q{2z \over b} \left(r \over w(z) \right)^2,
\label{phasgauss}
\end{equation}
where $w(z)=w_0\sqrt {1+4z^2/ b^2}$ and $w_0$ is the beam waist, related to the confocal parameter by $b=2\pi w_0^2/ \lambda$. The second contribution is the dipole phase, which depends on the intensity. The harmonics are here generated in the plateau cutoff transition region, where the dipole phase varies linearly with intensity, with a negative slope: $-\eta$. This contribution can then be written as: 
\begin{equation}
-\eta \times I(r,z)=-\eta {I_0\over 1+\left(2z\over b\right)^2} \exp \left(-2\left(r\over w(z)\right)^2\right) \simeq {-\eta I_0\over 1+\left(2z\over b\right)^2}+{2\eta I_0\over 1+\left(2z\over b\right)^2}\left(r\over w(z)\right)^2,
\label{phasdip}
\end{equation}
We can here assume $r<<w(z)$, since the harmonic profiles are much narrower than the fundamental.
In Chapter 3, we have considered the first terms of both contributions, that correspond to variations {\it on axis}. The second terms of these contributions, related to radial variations, both present a quadratic behavior. The term due to the focusing of the fundamental (Eq. (\ref{phasgauss})) is in fact the phase corresponding to a Gaussian harmonic field with the same confocal parameter. Its radial dependence is equal to $0.032r^2$ rad and is thus significantly slower than the ones observed for the harmonic fields. The reason for the bigger curvature of the phase fronts is thus the additional radial variation introduced by the dipole phase (second term in Eq. (\ref{phasdip})), which lies between $0.016r^2$ and $0.023r^2$ rad, depending on the considered intensity. 

\begin{figure}[h]
\vspace*{0.8cm}
\centerline{
\psfig{angle=90,height=6cm,file=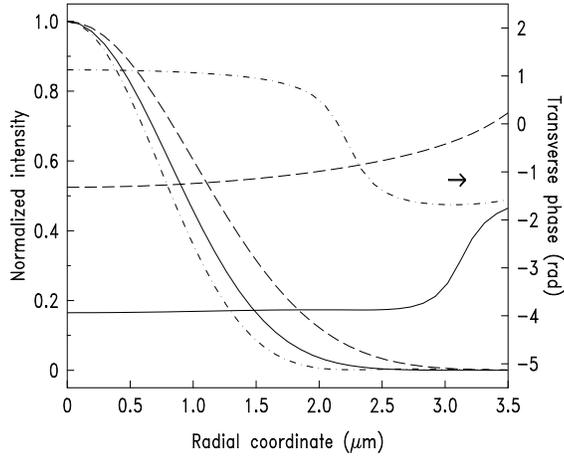}}
\caption{Radial variations of the intensity and phase of the harmonic field generated at the peak intensities: 4 $\times 10^{14}$ (dashed), 5 $\times 10^{14}$ (solid) and 6 $\times 10^{14}$ W/cm$^2$ (dot-dashed), and "backpropagated" to the focus position (3mm before the jet). 
\label{focus3}}
\vspace*{0.8cm}
\end{figure}

These curved phase fronts have important consequences, as we shall see below. The harmonic fields emitted at 4 $\times 10^{14}$ and 5 $\times 10^{14}$ W/cm$^2$ present Gaussian profiles and quadratic phases at the exit of the medium, and thus correspond to lowest order Gaussian modes, whose characteristics (beam waist position and confocal parameter) can be easily calculated. We find, in both cases, a virtual focus located at the laser focus (like in the perturbative case) but with an extremely small size: about 1.6 $\mu$m (even smaller than the perturbative 3.7$\mu$m)  The corresponding confocal parameters are also very small (about 1mm), so that the harmonic fields at the exit of the medium are already as if they were in the far-field: spherical phase front centered at the focus and linear increase of the beam size with the distance from the focus. Note that the radial variation of the phase of a spherical wave at a distance of $z=3.8$ mm is: $0.046r^2$ rad, close to the observed variations.

If we calculate the angular spread of these beams far from the jet, we find similar profiles, with half angle in $1/e^2$ between 3 and 5 mrad, that can be superimposed on the "near-field" profiles if their spatial dimension is calculated at a distance of $z=3.8$ mm. This is shown in dashed lines for the three intensities in Fig. \ref{nfprof3}, where the horizontal scale is interpreted as the divergence (half angle at half maximum) and is simply related to the radial coordinate by: divergence(mrad)=0.26 $\times r(\mu$m (or $r=3.8 \times$ divergence). Note that the profile corresponding to 6 $\times 10^{14}$ W/cm$^2$ is more distorted by the propagation, due to its multimode structure, but its divergence is still governed by its curved phase front at the exit of the medium.

\begin{figure}[h]
\vspace*{0.8cm}
\centerline{
\psfig{angle=0,height=9cm,file=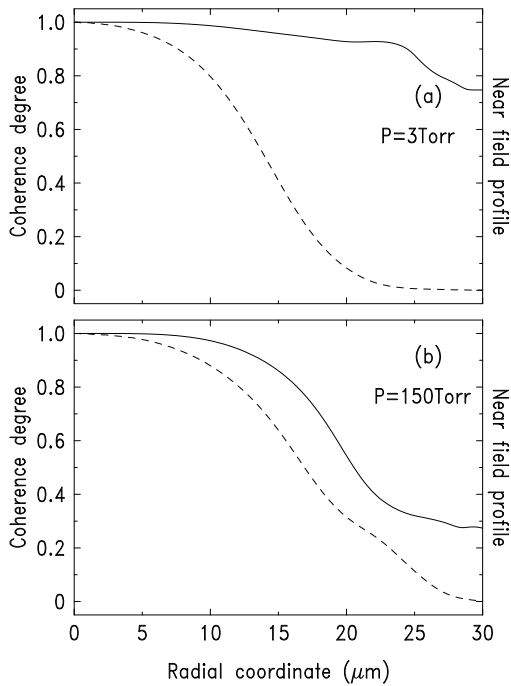}}
\vspace*{0.8cm}
\caption{Coherence degree (solid line) of the harmonic beam generated at $z=3$mm and at a peak intensity of 6 $\times 10^{14}$ W/cm$^2$, calculated at the exit of the medium between the central point and the outer points: for a peak pressure of (a) 3 Torr, (b) 150 Torr. The corresponding time integrated spatial profiles are shown in dashed lines. 
\label{cohdeg3}}
\end{figure}

Finally, we can check the preceding conclusions concerning the virtual source size, and extend them to the case 6 $\times 10^{14}$ W/cm$^2$, by "back propagating" these beams to their virtual focus. The solution of the propagation equations in free space for the complex conjugated field is formally equivalent to a time reversal, hence a "back propagation". Fig. \ref{focus3} presents the profiles and phases of the harmonic beams corresponding to the preceding intensities, calculated at the laser focus position ($z=0$). The phases stay about constant over the extent of the profiles, indicating quasi-plane phase fronts. This position is thus close to the best focus for these beams. The spot sizes are very small, close to the above estimates, with a minimum at 6 $\times 10^{14}$ W/cm$^2$ with a 1.4 $\mu$m radius at $1/e^2$. 

The very small size of the virtual source of the harmonic beams is a valuable information since it gives an indication on the size they can be refocused to (assuming ideal refocusing without aberrations introduced by the optics, and no reduction factor). From this, we can estimate the achievable {\it harmonic} intensities, in view of applications. If we take an harmonic pulse duration of 66 fs (see Section V) and an optimized number of photons of 10$^5$ (obtained by optimizing the pressure \cite{altucci}), the intensity at 18.3 nm (45th harmonic) reaches about 5 $\times 10^{8}$ W/cm$^2$. Note that if argon is used as generating medium instead of neon, a much higher conversion efficiency is obtained for lower order harmonics: e.g. 10$^9$ photons for the 19th harmonic \cite{suedois}. Since the behaviour of the dipole moment is very general, whatever the (sufficiently high) order or the generating gas, we can extend our results to the 19th harmonic in argon, and we thus find an harmonic intensity of 2 $\times 10^{12}$ W/cm$^2$ at 43.4 nm. If a reduction factor is introduced by the refocusing optics, the very nice harmonic phase front should allow one to reach an even smaller focus size, and thus a higher intensity, provided that the surface figure of the optics is good enough. These intensities have never before been reached at these short wavelengths, and would open the way to nonlinear optics in the XUV region.

\begin{figure}[h]
\vspace*{0.8cm}
\centerline{
\psfig{angle=90,height=6cm,file=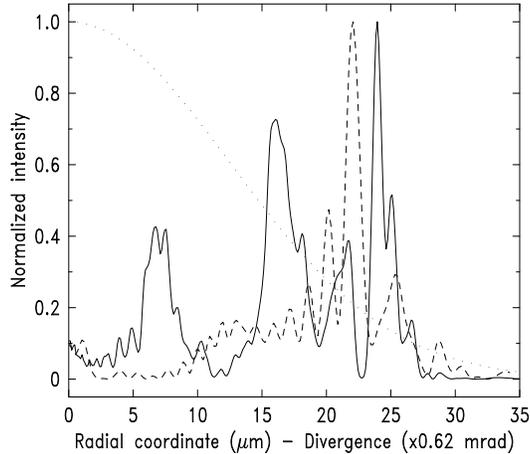}}
\vspace*{0.8cm}
\caption{Normalized spatial profile for the 45th harmonic at the maximum of the pulse for a jet position in $z=-1$mm and an intensity of 6 $\times 10^{14}$ W/cm$^2$. In solid line is shown the profile at the exit of the medium as a function of the radial coordinate, and in dashed line, the far-field profile (divergence). The fundamental near-field profile is presented in dotted line.
\label{nfprof1}}
\end{figure}

So far, we have studied the beam spatial characteristics at a given time during the laser pulse. Let us now consider the correlation {\it in time} of the fields inside the harmonic beam, by calculating the corresponding degree of spatial coherence, given by Eq.  (\ref{cohdeg}). It is shown in solid line in Fig. \ref{cohdeg3}(a), calculated at the exit of the medium ($z=3.8$ mm) between the central point (on the propagation axis) and the outer points, for a peak intensity of 6 $\times 10^{14}$ W/cm$^2$ and a peak pressure of 3 Torr. The coherence degree is very high, and stays above 0.9 over the whole extent of the time integrated spatial profile, which is shown in dashed line. At this low pressure, the influence of the free electrons generated by the (low) ionization of the medium is negligible. The small decrease of the coherence degree can be understood by the slow variation over the pulse of both the phase front curvature and the spatial profile, as shown for three different intensities in Figs. \ref{nfprof3} and \ref{nfphase3}. At a pressure of 15 Torr, the coherence degree is a little reduced, but stays very high (above 0.8). However, when the pressure is increased to 150 Torr, as shown in Fig. \ref{cohdeg3}(b), the spatial profile is broadened and somewhat distorted, while the coherence degree drops abruptly between 15 and 25 $\mu$m to a value of 0.3. 

Indeed, the free electron dispersion is not any more negligible and results in a phase shift that is spatially dependent, due to the spatial distribution of the free electron density. Close to the propagation axis, the laser intensity, and thus the free electron density, varies slowly radially, preserving the coherence. Further away, when the laser intensity drops, the phase shift imparted to the harmonic field is much smaller, resulting in a decorrelation with the center of the beam, and therefore, a smaller coherence degree. Ionization of the generating medium is thus an important cause of degradation of the coherence.

Here, we have calculated the coherence degree between the central point (on the axis) and the outer points. Note that, for two points taken symmetrically into the beam, the coherence degree is equal to 1, due to the revolution symmetry imposed to the problem. In experiments, this symmetry may be broken by different factors (gas jet and laser spot inhomogeneities etc...), which results in a smaller coherence degree, as recently measured by Ditmire {\it et al.} \cite{ditmire}. However, our study shows that, in order to truly characterize the coherence of the harmonic beam, it is necessary to measure the coherence degree for {\it non-symmetrical} positions. This point makes the situation very different from that of a beam originating from a completely incoherent source. In the latter case, the coherence degree only depends on the {\it relative} position of the two points considered, i.e. their distance, not on their {\it absolute} positions into the beam, as it is the case here. 

In conclusion for this focus position, at low pressure the very high coherence degree at the exit of the medium indicates that the harmonic beam is extremely coherent, in comparison to soft X-ray lasers \cite{Trebes,Celliers,xrays}. The coherence of soft X-ray lasers is often degraded by plasma fluctuations, resulting in a coherence degree equivalent to that of a spatially incoherent disk source with diameter of a few 100 $\mu$m \cite{Amendt}. The associated transverse coherence length at the output of the X-ray laser is a few $\mu$m, whereas it reaches several tens of $\mu$m for the harmonics.

\subsection{Study of the spatial coherence: atomic jet before the focus}

\begin{figure}[h]
\vspace*{0.8cm}
\centerline{
\psfig{angle=90,height=6cm,file=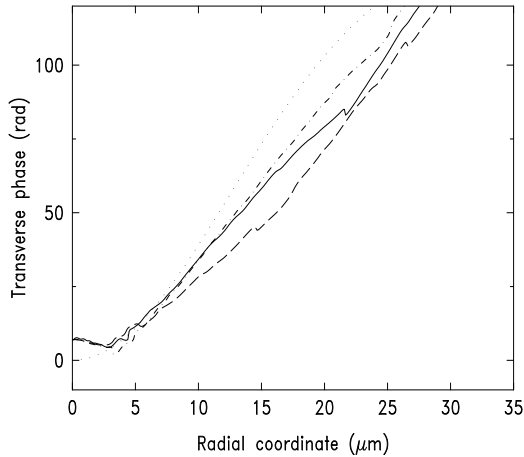}}
\vspace*{0.8cm}
\caption{Radial variation of the phases of the harmonic fields generated for a jet position in $z=-1$mm and intensities of respectively: 4 $\times 10^{14}$ (dashed), 5 $\times 10^{14}$ (solid) and 6 $\times 10^{14}$ W/cm$^2$ (dot-dashed). The radial variation of the polarization phase at the exit of the medium at an intensity of 6 $\times 10^{14}$ W/cm$^2$ is shown in dotted line.
\label{nfphase1}}
\end{figure}

The 45th harmonic spatial profile at the exit of the medium for a peak intensity of 6 $\times 10^{14}$ W/cm$^2$ and a focusing 1 mm after the jet is shown in solid line in Fig. \ref{nfprof1}. It is rather distorted and exhibits an annular structure with three major rings. Note that very little energy is emitted on axis. As shown in Chapter III, this is the result of the rapid variation of the polarization phase on axis, which favors the off axis phase matching. The external radius is 27 $\mu$m, larger than the 25 $\mu$m in $1/e^2$ of the fundamental beam (dotted line). The corresponding phase is presented in dot-dashed line in Fig. \ref{nfphase1} together with phases of beams generated at 4 $\times 10^{14}$ (dashed) and 5 $\times 10^{14}$ W/cm$^2$ (solid). Compared to the case $z=3$ mm, these phases are more irregular and vary much more quickly. Surprisingly they correspond to {\it diverging} phase fronts, even though the harmonics are generated by a {\it converging} beam. 

Consider the two contributions to the transverse phase of the polarization. The phase induced by the fundamental field, written in Eq.  (\ref{phasgauss}), is indeed negative but varies very slowly close to the focus (quasi-plane phase front). On the contrary, the contribution of the atomic phase, shown in Eq. (\ref{phasdip}), is associated to a very diverging phase front: not only the radial variation of the intensity is very rapid close to the focus, but also the harmonic is generated here in the plateau region where the average slope of the intensity dependence of the phase is twice as large as that in the cutoff region. Note that the quadratic approximation does not hold here, since we consider $r\approx w(z)$. The total polarization phase at the exit of the medium and for an intensity of 6 $\times 10^{14}$ W/cm$^2$ is shown in dotted line in Fig. \ref{nfphase1}. Its behavior is very similar to that of the phases of the harmonic beams, except for a slightly concave curvature. 

These very curved phase fronts dominate the propagation, and the calculation of the far field distribution of these beams gives annular profiles, as shown in dashed line in Fig. \ref{nfprof1}, where the horizontal scale is interpreted as the divergence and related to the radial coordinate by: $r=1.6 \times$ divergence. The external half-angle is 15 mrad, larger than the 10 mrad at $1/e^2$ of the fundamental. The harmonic is thus more divergent than the fundamental, which almost leads to a spatial separation of the two beams.

\begin{figure}[h]
\vspace*{0.8cm}
\centerline{
\psfig{angle=90,height=6cm,file=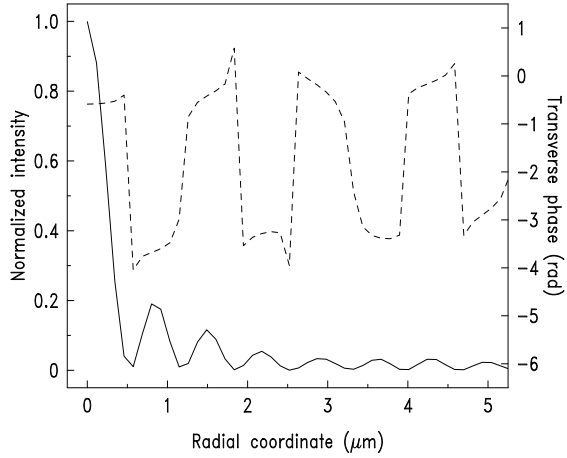}}
\vspace*{0.8cm}
\caption{Radial variations of the intensity (solid line) and phase (dashed line) of the harmonic field generated at $z=-1$mm and 6 $\times 10^{14}$ W/cm$^2$ and "backpropagated" to the entrance of the medium ($z=-1.8$mm). 
\label{focus1}}
\end{figure}

As before, we can "back propagate" this beam and find the position of the virtual focus. The highest harmonic intensity is obtained in $z=-1.8$ mm, which happens to be the entrance of the medium. The profile (solid line) and phase (dashed line) at this virtual focus are presented in Fig. \ref{focus1}. They look like the diffraction pattern of a ring with a spherical phase front. The phase shows a $\pi$ shift (i.e. a change of sign) for each profile oscillation, whereas the profile exhibits a narrow central peak (0.5 $\mu$m) surrounded by weak rings. A part of the energy is "lost" in these rings, so that the harmonic intensity reached at the central peak is barely larger than that obtained for a focusing 3 mm before the jet. 

\begin{figure}[h]
\vspace*{0.8cm}
\centerline{
\psfig{angle=90,height=6cm,file=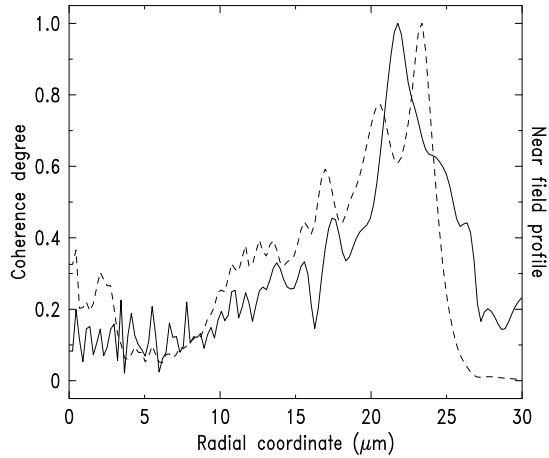}}
\vspace*{0.8cm}
\caption{Coherence degree (solid line) of the harmonic beam generated at $z=-1$mm and a peak intensity of 6 $\times 10^{14}$ W/cm$^2$, calculated at the exit of the medium between the point located close to the center of the ring ($r=22\mu$m) and the other points. The time integrated spatial profile is shown in dashed line. The peak pressure is 3 Torr.
\label{cohdeg1}}
\end{figure}

Let us now consider the coherence degree of this beam calculated at the exit of the medium for a peak pressure of 3 Torr, shown in solid line in Fig. \ref{cohdeg1}. Since there is very little energy emitted on axis (see the time-integrated profile, shown in dashed line), we take as a reference the point located at the center of the ring ($r=22$ $\mu$m). The coherence degree drops dramatically on both sides to about 0.3, and then oscillates around 0.2. The coherence of this beam is thus very much degraded, not by the free electron dispersion (low pressure), but by the rapid phase front fluctuations induced by the variation of the intensity in the laser pulse (see Fig. \ref{nfphase1}). The dynamically induced phase can thus be responsible for a dramatic degradation of the spatial coherence of the harmonic beam. Note that in traditional, i.e. perturbative, low-order harmonic generation, where the harmonic dipole moment does not exhibit an intrinsic phase and depends regularly on the laser intensity, the coherence of the laser beam is simply transmitted to the harmonic beam. In conclusion, the dynamically induced phase plays a central role in determining the spatial characteristics of the generated harmonic beams, and especially their coherence properties, which are very sensitive to the focusing conditions.

\section{Temporal and spectral coherence}

\subsection{Influence of the jet position}

\begin{figure}[h]
\vspace*{0.8cm}
\centerline{
\psfig{angle=0,height=9cm,file=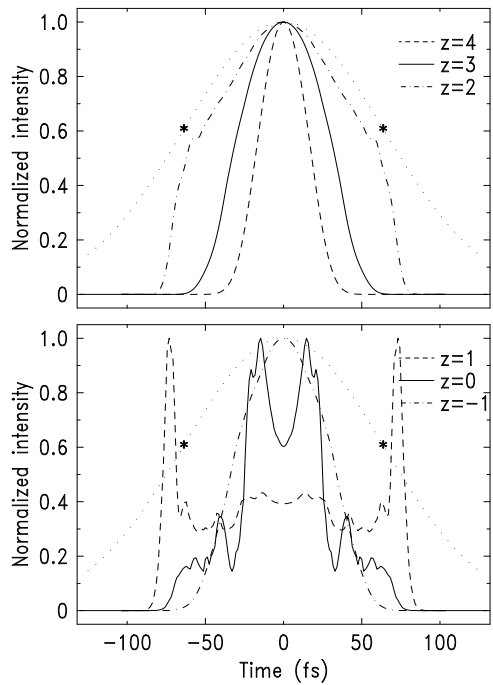}}
\vspace*{0.8cm}
\caption{Temporal profiles of the 45th harmonic generated at 6 $\times 10^{14}$ W/cm$^2$, for different $z$ positions indicated in mm in the captions. The laser profile is presented in dotted lines, the stars indicating the position of the inflection points.
\label{temp}}
\end{figure}

We have shown in the preceding chapter that the dynamically induced phase results in a dynamic phase matching due to the intensity distribution in the laser pulse. This will affect the temporal as well as the spectral properties of the generated harmonics. Let us first consider the harmonic temporal profile. Since we calculate in the slowly varying envelope approximation, the temporal profile is simply obtained from the response to elemental square laser pulses, with a Gaussian distribution of the intensities. Fig. \ref{convstat} shows clearly that the temporal behavior will be very much dependent on the position of the jet. On both sides of the conversion efficiency curve, for example in $z=-1$ mm and in $z=3$ mm, the efficiency increases quickly with intensity. The harmonic yield will thus be  high mainly for intensities close to the maximum of the laser temporal envelope, leading to narrow and regular harmonic  
temporal profiles. On the contrary, for intermediate $z$ positions, phase matching can be more efficient for low intensities than for the maximum of the laser envelope, resulting in large and distorted harmonic profiles.

These predictions are confirmed in Fig. \ref{temp}, which presents the 45th harmonic temporal profiles obtained at 6 $\times 10^{14}$ W/cm$^2$ for different $z$ positions (ionization is not yet taken into account). When the jet is moved from $z=4$ mm to $z=2$ mm, the harmonic profile gets larger (from 36 to 125 fs FWHM) but remains regular. On the other side, in $z=-1$ mm, the profile is also smooth, narrow and very similar to the one obtained in $z=3$ mm, with the same FWHM (67 fs). On the contrary, in $z=1$ mm and $z=0$, the profiles are quite distorted, with fast fluctuations. In $z=1$ mm, phase matching is much more efficient at 3 $\times 10^{14}$ W/cm$^2$ than at higher intensities (cf Fig. \ref{convstat}), which leads to narrow peaks in the harmonic profile at half maximum of the laser envelope (shown in dotted lines). Rapid fluctuations also appear in the profile corresponding to the position $z=0$. They are probably due to the oscillations of the slope in the intensity dependence of the dipole phase in the plateau region, that induce rapid changes of the dynamic phase matching. 


\begin{figure}[h]
\vspace*{0.8cm}
\centerline{
\psfig{angle=0,height=9cm,file=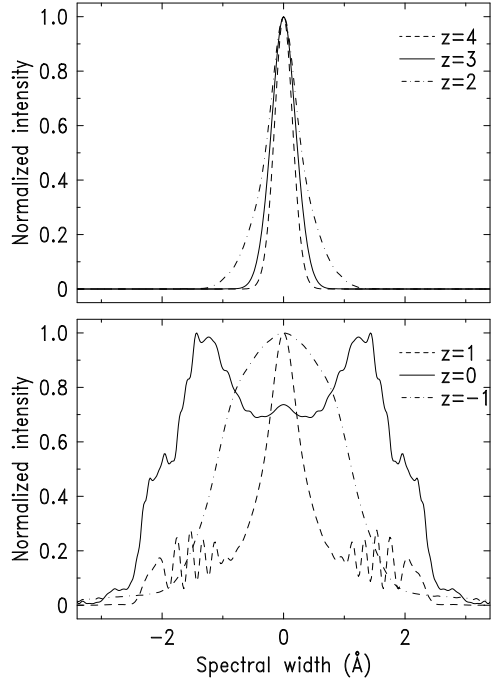}}
\vspace*{0.8cm}
\caption{Spectral profiles of the 45th harmonic generated at 6 $\times 10^{14}$ W/cm$^2$, for different $z$ positions indicated in mm in the captions.
\label{spec}}
\end{figure}

The corresponding spectral profiles are presented in Fig. \ref{spec}. For a focusing sufficiently before the medium, the spectra are quite narrow (e.g. 0.4 \AA \ FWHM at $z=3$ mm). When the jet is located at $z=1$ mm, the base of the profile gets broadened with superimposed oscillations. In $z=0$, the profile is very broad with a FWHM of 4 \AA.  Finally, in $z=-1$ mm, the profile is again regular with a width of 2.2 \AA. These large variations in spectral width cannot be explained by variations in the intensity temporal profiles. Note for example that in $z=-1$ and $z=3$ mm, the temporal profiles are very similar with the same FWHM, whereas there is a factor of 5 between the corresponding spectral widths. Consequently, these broad spectra are not due to variations of the {\it amplitude} of the harmonic fields, but to variations of their {\it phase}. 
Calculations performed without taking into account their phase, i.e. assuming  
that the fields are Fourier transform limited, give indeed much narrower spectra, with widths smaller than 0.15 \AA.

 The origin of this phase modulation is the temporal variation of the dipole phase during the laser pulse. Let us model the intensity dependence of the dipole phase by a linear decrease, whose slope $-\eta$ depends on the region of the spectrum: in the cutoff $\eta=13.7 \times 10^{14}$ rad/(W/cm$^2$) and in the plateau $\eta=24.8 \times 10^{14}$ rad/(W/cm$^2$). The induced modulation of the emitted harmonic field reads: 
\begin{equation}
\Delta \Phi (t)=-\eta I(t)=-\eta I_0 \exp (-4\ln 2(t/\tau)^2)
\label{phasmod}
\end{equation}
where $\tau$ is the FWHM of the Gaussian laser envelope. This results in a modulation of the instantaneous frequency of:
\begin{equation}
\Delta \omega (t)=-{\partial (\Delta \Phi)\over \partial t}=\eta {\partial I(t)\over \partial t}
\label{freqmod}
\end{equation}
with a corresponding broadening of the spectrum. This phenomenon is thus very similar to self-phase modulation of an intense laser pulse in a Kerr medium with a negative index $n_2$ \cite{Shen}. However, the phase modulation is here induced on the harmonic, whose temporal profile is different from that of the laser, and we shall see the consequences below.

The rising edge of the harmonic pulse is thus shifted to the blue, and the falling edge, to the red. The extremal instantaneous frequencies are symmetrical on either side of the central frequency, and correspond to the inflection points of the laser temporal envelope: $t_i=\pm \tau /\sqrt{8ln2}$. One then finds:
\begin{equation}
\Delta \lambda _{ext}={-\lambda ^2 \over 2 \pi c}\Delta \omega _{ext}=\pm {\lambda ^2 \over 2 \pi c}{\sqrt{8\ln 2} \over \tau}\exp (-1/2)\eta I_0
\label{lambmod}
\end{equation}
and in our conditions: $\Delta \lambda _{ext} \simeq \pm 1.6 \times 10^{-2} \eta I_0$ (\AA). The maximal broadening of the spectrum should thus be observed at focus, which experiences the largest intensity variation ($I_0=6 \times 10^{14}$ W/cm$^2$) and where the intensity at the inflection points is still in the plateau region, ensuring the larger slope $\eta$. This results in $\Delta \lambda _{ext} \simeq \pm 2.4$ \AA, close to the broadening observed for $z=0$. However, this spectrum does not exhibit the characteristic grooved shape of the self phase modulation spectra. 

Let us recall the main features of self phase modulation \cite{Shen}. Close to the inflection points, the instantaneous frequency varies slowly, unlike in the vicinity of the pulse maximum. The spectral density is thus maximum on the sides (in $\Delta \lambda _{ext}$) and minimal at the central frequency.  On the $\Delta \Phi (t)$ curve, there exist two points of the same slope, located on either side of the inflection point. They thus correspond to the same instantaneous frequency and interfere constructively or destructively depending on their relative phase. This results in a grooved spectrum, with clear peaks and valleys. 
In our case, the phase modulation is induced on the harmonic, which is generated efficiently for large enough intensities, located in general above the inflection points (cf Fig. \ref{temp}). Consequently there will be no interferences in the spectra. They thus are more regular, except the one in $z=1$ mm which corresponds to the only temporal profile sufficiently large to go past the inflection points.  

\begin{figure}[h]
\vspace*{0.8cm}
\centerline{
\psfig{angle=0,height=9cm,file=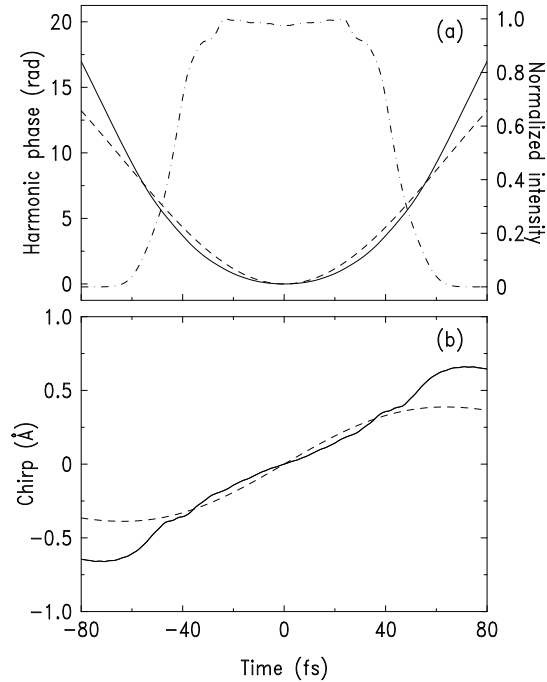}}
\vspace*{0.8cm}
\caption{(a) Temporal dependence of the intensity (dot-dashed) and phase (solid line) of the harmonic field exiting the medium on axis ($r=0$) for a focusing 3 mm before the jet. The variation of the polarization phase at that point is shown in dashed line. (b) Corresponding frequency chirps for the field (solid) and the polarization (dashed).
\label{mod3}}
\end{figure}

In the following, we study in more details what happens in $z=3$ and $z=-1$ mm. In the case $z=3$ mm, the integrated spatial profile at the exit of the medium is very regular, with a maximum on axis. The phase temporal variation of the harmonic field in $r=0$ has thus a large incidence on the full spectrum, and is characteristic of the phenomenon. Fig. \ref{mod3}(a) presents the temporal variation of the intensity (dot-dashed line) and phase (solid line) of the harmonic field emitted on axis at the exit of the medium. The intensity profile is quasi square, with a FWHM of 90 fs. The phase presents a very regular behavior. Let us compare it with the polarization phase (shown in dashed line) calculated at this point from Eq. (\ref{phasmod}). The maximum intensity there is $I_0=1.8 \times 10^{14}$ W/cm$^2$ (the exit of the medium is 3.8 mm away from the focus), which corresponds to the cutoff region (small value of $\eta$). The two curves are very similar over the width of the intensity profile, confirming the origin of the phase modulation. The induced frequency chirps are presented in \AA \ in Fig. \ref{mod3}(b). The spectral width corresponding to efficient harmonic emission is about 0.5 \AA, close to that of the full spectrum. Note that the chirp is almost linear during the harmonic pulse.

\begin{figure}[h]
\vspace*{0.8cm}
\centerline{
\psfig{angle=0,height=9cm,file=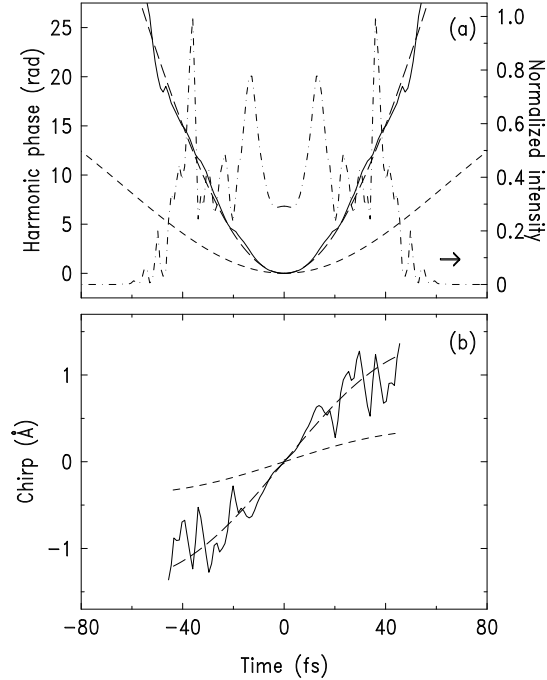}}
\vspace*{0.8cm}
\caption{(a) Temporal dependence of the intensity (dot-dashed) and phase (solid line) of the harmonic field exiting the medium in $r=20 \mu$m for a focusing 1 mm after the jet. The variation of the polarization phase at that point is shown in short-dashed line, and in a point of maximum intensity of 3.4 $\times 10^{14}$ W/cm$^2$, in long-dashed line. (b) Corresponding frequency chirps for the field (solid) and the polarization (long and short-dashed).
\label{mod1}}
\end{figure}

In $z=-1$ mm, the near-field spatial profile is annular due to an efficient phase matching off axis (cf Chapter 4). Let us consider the harmonic field in $r=20 \mu$m at the exit of the medium, where the emission is efficient. The intensity profile, shown in dot-dashed line in Fig. \ref{mod1}(a), exhibits rapid oscillations superimposed on a 90 fs wide curve. These fluctuations do not appear in the spatially integrated temporal profile of Fig. \ref{temp}, which is very regular. This implies that they are compensated by opposed oscillations of the emission in neighboring points. This can be understood by the variation of the phase front curvature in time, which leads to fluctuations of the angle of the emission cone. 

In a given position, one then sees the radiation going back and forth, hence these oscillations. The phase of the harmonic field, shown in solid line, varies much more rapidly than in $z=3$ mm. Since the position considered is far from the axis, the maximum intensity corresponds to the cutoff region : $I_0=1.7 \times 10^{14}$ W/cm$^2$, and the polarization phase at that point (short-dashed line) varies too slowly to explain the harmonic phase behavior. A good agreement is obtained if we take an intensity of $I_0=3.4 \times 10^{14}$ W/cm$^2$, with the plateau slope $\eta$ (long-dashed line). This implies that the harmonic radiation which exits the medium at that point has mainly been generated in a deeper region, close to the axis. The phase front curvature makes it diverge and exit at that position. 
The frequency chirp, presented in Fig. \ref{mod1}(b), shows the phase sudden changes of slope, reflecting the intensity fluctuations. The corresponding spectral width is about 2.2 \AA, close to that of the full spectrum.

The origin of the difference by a factor 5 of the spectral widths in $z=3$ and $z=-1$ mm appears thus clearly: in $z=3$ mm, the harmonic emission corresponds to the cutoff region, hence a small slope $\eta$ and a low intensity. In $z=-1$ mm, it is mainly in the plateau, with a slope twice as large, and for intensities at least two times larger. The spatial and spectral coherence properties are thus closely linked, and are both governed by the variations of the atomic phase. 

\subsection{Influence of the ionization}

\begin{figure}[h]
\vspace*{0.8cm}
\centerline{
\psfig{angle=0,height=9cm,file=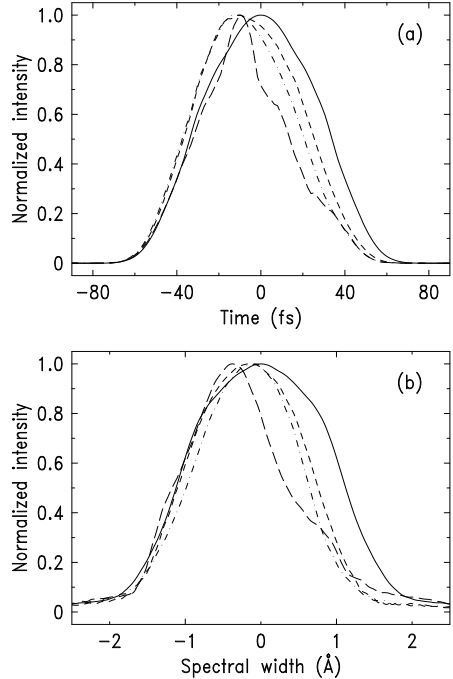}}
\vspace*{0.8cm}
\caption{Temporal (a) and spectral (b) profiles for a focusing 1mm after the jet at 6 $\times 10^{14}$ W/cm$^2$: without ionization (solid), with ionization at a pressure of 3 Torr (short-dashed) and 15 Torr (long-dashed) of neon. In dot-dashed line is shown the profile corresponding to 15 Torr when the defocusing of the laser is neglected.
\label{ioni}}
\end{figure}

When the jet is placed several mm away from the focus, the intensity in the medium is sufficiently low to neglect the effects of ionization. In $z=3$ mm for example, the profiles are only slightly modified. We will thus concentrate on the case $z=-1$ mm, and try to separate the different effects: depletion of the atomic population, defocusing of the laser and phase mismatch both due to the generated free electrons.

Fig. \ref{ioni}(a) presents the distortions of the temporal profile induced by the different effects. In short-dashed line is shown the profile corresponding to a peak pressure of 3 Torr. This low pressure ensures that the free electron density is low enough to neglect its effects. This profile thus characterizes the incidence of the depletion of the medium on harmonic generation (the susceptibility of the generated ions is supposed to be negligible compared to the atoms). At the center of the jet, ionization reaches 67\% at the end of the laser pulse. The resulting profile is clearly asymmetric, with a rising edge  hardly modified and a decrease of the efficiency on the falling edge. The FWHM is slightly decreased, from 67 to 63 fs. 

At a pressure of 15 Torr, the effects of the free electrons are not any more negligible. We have first made the calculations by neglecting the defocusing of the laser: we suppose a Gaussian propagation and we only solve the propagation equation for the harmonic field in the ionized medium. The temporal profile (dot-dashed) is even more distorted on the falling edge: the width decreases to 58 fs. This is due to the phase mismatch induced by the free electron dispersion, that reaches $\Delta k \simeq 22$ mm$^{-1}$ at the center of the jet and at the maximum of the pulse.

The study of the propagation of the laser in the ionizing medium shows clearly an effect of defocusing: at the exit of the medium (0.2mm from the focus), the maximum intensity is reduced by 17\%, to 5 $\times 10^{14}$ W/cm$^2$. The reduced intensity experienced by the medium results in a smaller ionization degree: 54\% at the center of the jet and at the end of the pulse (vs 67\%). A consequence of the defocusing is thus to reduce the amplitude of the preceding effects: less depletion and smaller free electron density. However, the corresponding temporal profile (long-dashed) is the most distorted one. The efficiency drops abruptly after the maximum of the pulse, leading to a width of 46 fs. Two elements can explain this decrease in efficiency. First, the reduced intensity induces a decrease of the polarization amplitude. Second, the defocusing changes the intensity distribution in the medium compared to the Gaussian. The atomic phase distribution is thus modified, which may result in a less efficient phase matching.

In Fig. \ref{ioni}(b) we show the incidence of these phenomena on the spectrum. A striking feature is that, whatever the effect considered, the distortions of the spectrum simply follow the distortions of the temporal profile. The blue side of the spectrum is weakly affected, whereas its red side is progressively cut off. The FWHM decreases first from 2.2 to 1.9 \AA \  due to the depletion (3 Torr), then to 1.6  \AA \ due to the phase mismatch induced by the free electrons at 15 Torr, and finally to 1.5 \AA \ when the defocusing of the laser is taken into account. This is a clear illustration of the phase modulation phenomenon. Since the rising edge of the harmonic pulse is weakly affected by ionization, it is also the case for the blue side of the spectrum it is associated to. On the contrary the falling edge, and consequently the red side of the spectrum, are strongly distorted. This results in a blueshift of the central frequency of the spectrum. Note that this phenomenon is different from the blueshift induced by the temporal variation of the free electron density, and thus of the refractive index \cite{suedois,rae}. In our conditions, the latter effect shifts the fundamental spectrum by 2.4 \AA, which results for the 45th harmonic in a shift of only 0.05 \AA.

\subsection{Consequences of the phase modulation}

\begin{figure}[h]
\vspace*{0.8cm}
\centerline{
\psfig{angle=0,height=9cm,file=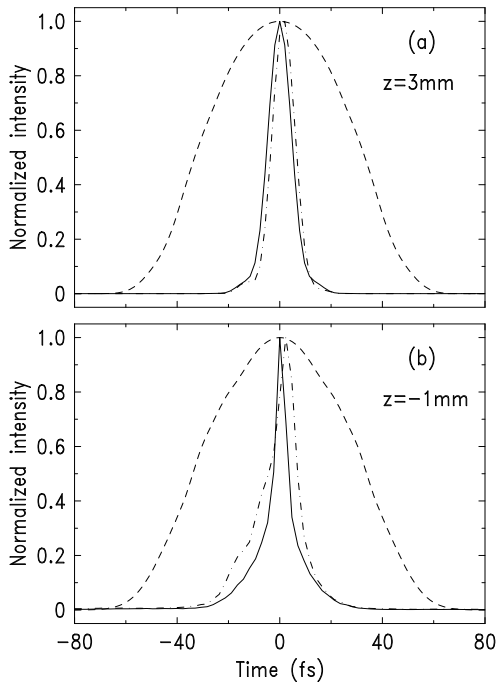}}
\vspace*{0.8cm}
\caption{Compressed temporal profiles obtained at 6 $\times 10^{14}$ W/cm$^2$ and at a pressure of 3 Torr (solid) and 15 Torr (dot-dashed) of neon: (a) in $z=3$mm, (b) in $z=-1$mm.  In dashed lines are shown the corresponding emission profiles before compression when ionization is neglected.
\label{compress}}
\end{figure}

The phase modulation studied below can be used to control the temporal and spectral properties of the harmonic radiation. In the time domain, the regular close-to-quadratic phase, and the corresponding linear chirp, lead to the possibility of compressing the harmonic pulse with a pair of gratings, as is done in CPA lasers. Schafer and Kulander studied the single atom response to very short laser pulses (27 fs), and showed that it should be possible to compress the harmonic pulse to durations below the fundamental period (2.7 fs) \cite{kenpriv}. We will discuss the short duration problem in the next section, and we concentrate here on the study of the compression of the {\it macroscopic} harmonic beam.

To simulate the compression, we subtracted a mean quadratic phase from the phase of the spectrum emitted at each point at the exit of the medium. The inverse Fourier transform followed by the spatial averaging give the compressed macroscopic temporal profile. The result of the compression in $z=3$ mm for a peak pressure of 3 Torr is presented in solid line in Fig. \ref{compress}(a). The profile is very narrow, with a FWHM of 10 fs (vs. 66 fs before compression, dashed line). This is close to the Fourier transform limit. The short duration is maintained when the pressure is increased to 15 Torr (dot-dashed line). In $z=-1$ mm, the compressed profile for a pressure of 3 Torr is even narrower, as shown in solid line in Fig. \ref{compress}(b): 6 fs vs. 67 fs before compression. 

Note that on the basis of the spectral width which is more than three times larger than in $z=3$ mm, we would have expected a smaller duration. However, this optimal compression is not reached due to temporal fluctuations in the phase (see Fig. \ref{mod1}) and to spatial inhomogeneities of the phase modulation. Phase matching has thus to be considered carefully in order to get the shortest harmonic pulses. At a pressure of 15 Torr, not only is the spectral width reduced (due to defocusing of the laser, see preceding section), but also the free electron dispersion introduces a temporally-varying phase shift on the harmonic emission. This results in a broader temporal profile of 13 fs FWHM, shown in dot-dashed line. In conclusion, we have shown that it is possible to compress the {\it macroscopic} harmonic beam generated by a 150 fs laser to durations below 10 fs.

\begin{figure}[h]
\vspace*{0.8cm}
\centerline{
\psfig{angle=0,height=9cm,file=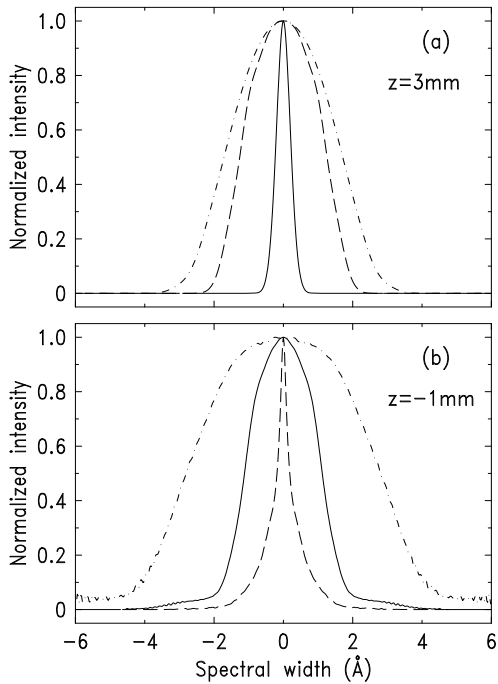}}
\vspace*{0.8cm}
\caption{Spectral profiles of the 45th harmonic generated with a fundamental without chirp (solid), chirped positively (dashed) or negatively (dot-dashed): (a) in $z=3$mm, (b) in $z=-1$mm. Ionization is not taken into account.
\label{chirp}}
\end{figure}

In the frequency domain, it should be possible to compensate for (or increase) the dynamically induced harmonic chirp by introducing an appropriately designed frequency chirp on the fundamental beam. We have calculated the harmonic spectrum generated by a laser pulse of same duration (150 fs) but presenting a quadratic phase in time, such that the fundamental spectrum is broadened to 32 nm, corresponding to a Fourier transform limit of 25 fs. Fig. \ref{chirp}(a) presents the spectra obtained for positive and negative chirps in the case of a focusing 3 mm before the jet. In the preceding section, we have seen that for this position, the dynamically induced harmonic chirp, which is negative, is limited (cf Fig. \ref{mod3}), leading to a quite narrow spectrum (0.4 \AA). It is much smaller than the chirp induced by the fundamental on the harmonic (recall that the fundamental phase is multiplied by the nonlinear order in the polarization). In these conditions, we thus get a broadening of the harmonic spectrum whatever the sign of the fundamental chirp. 

However, when the fundamental chirp is negative (blue before the red), it adds to the negative dynamic chirp, leading to a 3.4 \AA \ width (dot-dashed line). On the other hand, the positive fundamental chirp gets subtracted from the dynamic chirp, giving a 2.5 \AA \ width (dashed line). 
Note that the fundamental chirp could be adjusted to reduce the width of the spectrum for this particular position. In $z=-1$ mm, the dynamic harmonic chirp is larger resulting in a 2.2 \AA \ wide spectrum (solid line in Fig. \ref{chirp}(b)). It is in fact of the same order as the one induced by the fundamental chirp. When both add (negative fundamental chirp), this results in an extremely wide spectrum (5.6 \AA, dot-dashed line). Conversely, when they get subtracted (positive fundamental chirp), they compensate each other, leading to a very narrow spectrum of 0.4 \AA \ width (dashed line).

\begin{figure}[h]
\vspace*{0.8cm}
\centerline{
\psfig{angle=90,height=6cm,file=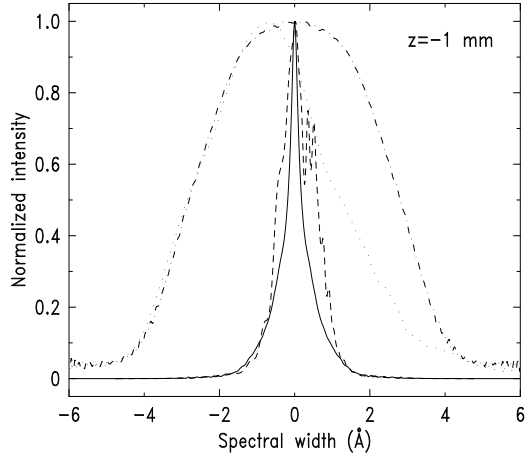}}
\vspace*{0.8cm}
\caption{Spectral profiles of the 45th harmonic generated in 15 Torr of neon at 6 $\times 10^{14}$ W/cm$^2$, in $z=-1$mm and with a fundamental chirped positively with (resp. without) ionization: short-dashed (solid), or negatively with (resp. without) ionization:  dotted (dot-dashed).
\label{chirpion}}
\end{figure}

So far, we have not yet taken ionization into account, which may modify the results in $z=-1$ mm at a sufficiently high pressure. At a pressure of 3 Torr, only depletion plays a role, and its influence is quite limited (see Fig. \ref{ioni}). The consequences of ionization for a pressure of 15 Torr are shown in Fig. \ref{chirpion}. The wide spectrum (negative chirp) is reduced to 4.0 \AA \ (dotted line) by the same effect that was described in the preceding section: loss of the red part of the spectrum due to a reduced efficiency on the falling edge of the pulse. On the contrary, the ``compensated" spectrum (positive chirp) is broadened to 1.1 \AA \ (short-dashed). This is certainly caused by phase fluctuations induced by the free electron dispersion. 

In conclusion, the regular behavior of the dynamically induced harmonic chirp can be used to control the temporal as well as the spectral properties of the harmonic emission. Concerning the latter, we find a qualitative agreement with the experimental results of Zhou {\it et al.} \cite{kapteyn}: for negative fundamental chirps, they observe a very pronounced broadening whereas for positive chirps, the peaks remain relatively narrow. However, the experiments were performed in argon at a relatively high intensity, thus with a higher degree of ionization than considered here. Harmonic generation occurs then only on the rising edge of the laser pulse, resulting in significant redshifting or blueshifting (depending on the fundamental chirp) of the harmonic radiation.

\subsection{Influence of nonadiabatic phenomena}

So far, we have considered harmonic generation by laser pulses of 150 fs duration. The recent advent of lasers of duration 25 fs and less, sufficiently intense to generate harmonics, motivates the investigation of shorter pulse effects. Except for the influence of ionization which depends directly on time, our results are simply scalable to shorter durations as long as the approximations made still apply: for example, going from 150 to 25 fs laser pulses (while keeping the same peak intensity) would result in a shortening of a factor 6 of the harmonic temporal profile and a corresponding broadening of the same factor of the spectrum. What about our approximations? 

The slowly varying envelope approximation {\it for the propagation} of the harmonic field is valid for very short laser pulses since the harmonic period is much less than 1 fs. However, the very broad spectra of the generated harmonics may overlap and interfere, making the propagation calculations more difficult. But the main problem is to take into account the nonadiabatic response of the nonlinear polarization to the fast driving field. In our calculations, we consider that the harmonic dipole moment follows adiabatically (or reacts instantaneously to) the intensity distribution of the laser pulse. In particular, we consider that the laser field amplitude does not vary significantly over the optical period. For the very short pulses mentionned above, this may not be true and result in a distorted atomic response. 

Such a nonadiabatic effect was invoked by Schafer and Kulander to explain the phase characteristics of harmonics generated by a 27 fs laser pulse \cite{kenpriv}. They calculated the single argon atom response by integrating numerically the time dependent Schr\"odinger equation in the single active electron approximation. The plateau region of the spectrum is highly structured, the expected odd harmonics being hard to distinguish. However, harmonics at the end of the plateau are found very broad but distinct, exhibiting a nice quadratic phase. They interprete this phenomenon in terms of the change in laser intensity during the laser period that alters the harmonic generation process \cite{Watson}. An electron that enters the continuum while the laser intensity is increasing experiences an additional acceleration before returning. This produces a blueshift on the rising edge of the laser envelope. Conversely, electrons ionized after the peak of the pulse are decelerated, returning later, leading to a redshift of the spectrum. 

\begin{figure}[h]
\vspace*{0.8cm}
\centerline{
\psfig{angle=90,height=6cm,file=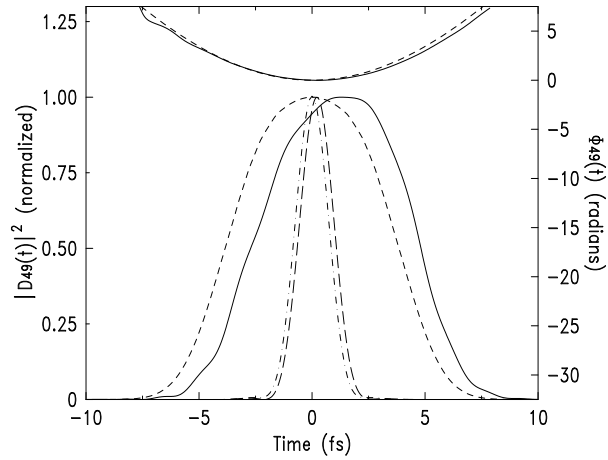}}
\vspace*{0.8cm}
\caption{Temporal envelope (lower curves) and phase (upper curves) of the 49th harmonic generated in argon by a 27 fs laser pulse at 3 $\times 10^{14}$ W/cm$^2$: adiabatic (short-dashed) and nonadiabatic (solid) calculations. The temporal profiles after compression are shown in dot-dashed (adiabatic) and long-dashed (nonadiabatic).
\label{tempadia}}
\end{figure}

\begin{figure}[h]
\vspace*{0.8cm}
\centerline{
\psfig{angle=90,height=6cm,file=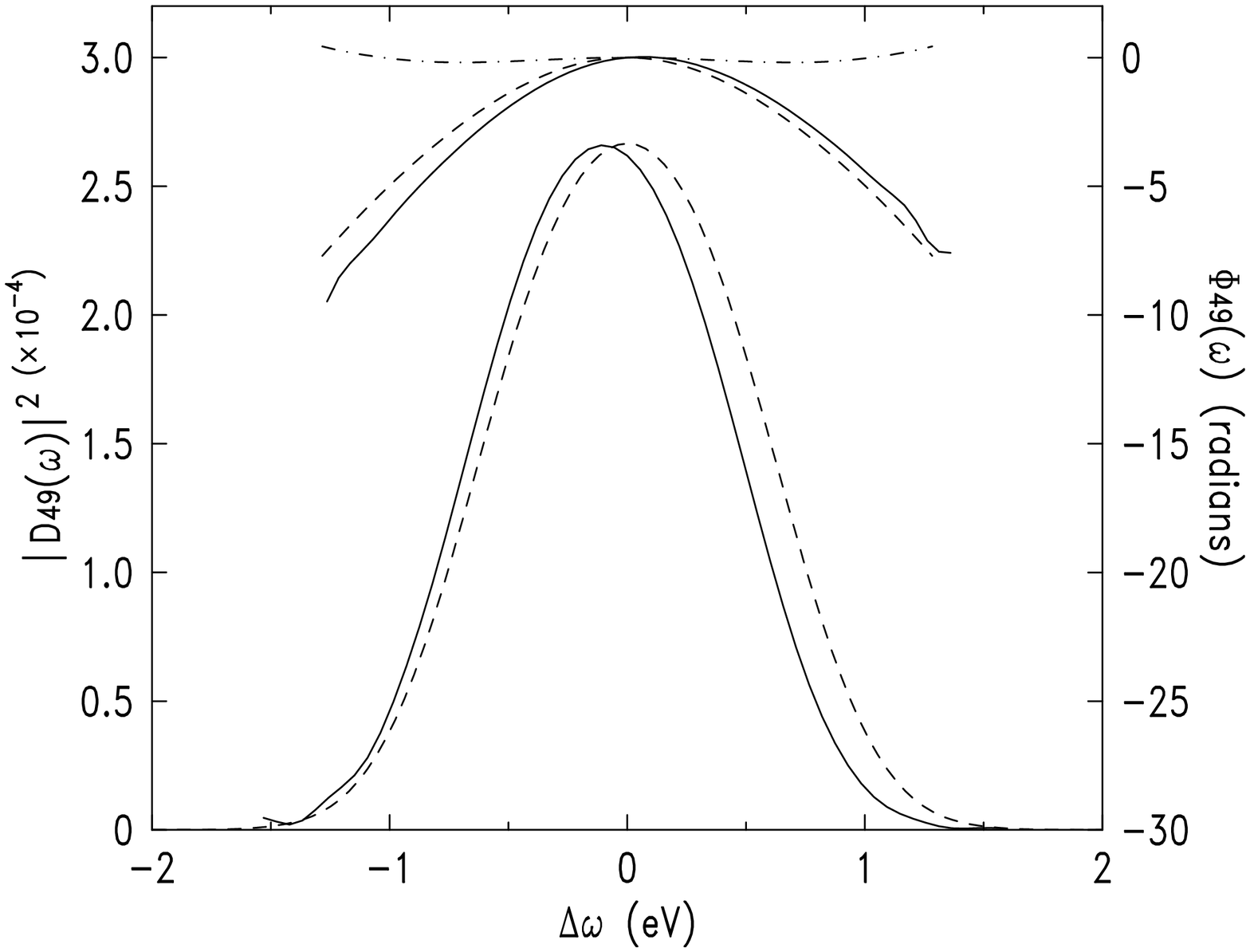}}
\vspace*{0.8cm}
\caption{Spectral envelope (lower curves) and phase (upper curves) of the 49th harmonic generated in argon by a 27 fs laser pulse at 3 $\times 10^{14}$ W/cm$^2$: adiabatic (short-dashed) and nonadiabatic (solid) calculations. The adiabatic phase minus its quadratic component is shown in dot-dashed line.
\label{specadia}}
\end{figure}

In order to investigate these effects, we calculated the nonadiabatic single atom response to the fast driving field in the strong field approximation and compared it with the adiabatic result. We use the same conditions as in Ref. \cite{kenpriv}, i.e. an argon atom interacting with a 27 fs, 810 nm pulse with a peak intensity of 3 $\times 10^{14}$ W/cm$^2$. In these calculations, we neglect ionization since the considered intensity is (slightly) below saturation for this small pulse duration.

The 49th harmonic temporal envelopes and phases are presented in Fig. \ref{tempadia}. The adiabatic profile (short-dashed) is very regular with a 7.6 fs FWHM, while the corresponding phase (upper short-dashed) exhibits a close-to-quadratic temporal dependence. Indeed, at the considered intensity, the 49th harmonic is in the cutoff region, where the dipole varies regularly with intensity. In particular, the phase depends linearly with intensity, which leads to a temporal quadratic phase close to the top of the Gaussian laser pulse. The nonadiabatic profile (solid) exhibits the same width but is delayed compared to the adiabatic by roughly 1.3 fs, whereas the phase is close to the adiabatic calculation. This shift in time can be related to the physics of the process. In the two-step model, there is one main electron trajectory for the harmonic emission in the cutoff region. The corresponding return time, i.e. the time between tunneling and recombination, is about half a period. Since the amplitude of the emission is mainly determined by the ionization probability,   
this explains the delayed response of the atomic dipole.

The spectral envelopes and phases are presented in Fig. \ref{specadia}. The adiabatic profile is very broad (1.3 eV FWHM), even more than the nonadiabatic (1.2 eV). The latter is shifted to the red, which is a consequence of the shift in time of the corresponding temporal profile. Indeed, the blue spectral components are associated to the decreasing part of the temporal quadratic phase, and are thus affected by the delay of the rising edge of the temporal envelope. Furthermore this positive temporal quadratic phase results in a negative spectral quadratic phase, as shown in Fig. \ref{specadia} (upper lines). The adiabatic phase minus its quadratic component, shown in dot-dashed line, is almost constant. This operation can be performed experimentally with a pair of gratings, leading to the compression of the harmonic temporal profile. They are presented in Fig. \ref{tempadia}. The adiabatic (dot-dashed) and nonadiabatic (long-dashed) profiles are similar with a 1.7 fs FWHM. The 
compression by more than a factor 4 results in a corresponding increase in peak intensity. All these profiles are very similar to the ones reported in Ref. \cite{kenpriv}. 

In conclusion, the main features in the cutoff region such as broad spectrum and quadratic phase are very well reproduced by an adiabatic calculation. They are thus the consequence of the intensity dependence of the phase of the adiabatic dipole, and not of a nonadiabatic effect due to the change in intensity during the laser period. In fact, harmonics in the cutoff are created close to the peak of the laser pulse, where the intensity varies slowly. Nonadiabatic phenomena may play a role in the plateau region, where the harmonics are also generated on the edges of the laser envelope. In the cutoff, the main nonadiabatic effect seems thus to be the shift in time of the atomic response. Note that if we take a sine laser field instead of a cosine, the harmonic temporal profile is slightly modified, exhibiting a double peak structure for the same width. This is an indication that we are close to the limit of validity
of the adiabatic approximation. 

Another important element is the width of the  
 spectrum. It spans one laser photon energy (1.5 eV) on either side of the harmonic frequency. This means that for shorter pulse durations, the spectra of adjacent harmonics will overlap and interfere, leading to a perturbed behaviour also in the cutoff region. Note that a nonadiabatic effect has been invoked by Christov {\it et al.} to explain the delayed ionization with short laser pulses (25 fs) and the corresponding increase in harmonic plateau extension \cite{Christov}.

\section{Future applications}

High order harmonics are currently used in a number of applications such as atomic and molecular spectroscopy, solid state physics... (for a review, see Ref. \cite{application}). We focus here on potential applications involving their coherence properties. Indeed, the main message of the studies presented in Chapters 3-5 is that harmonics can be generated in the form of short, but spatially and temporally coherent pulses. Moreover, the properties of such pulses can be to a great extent and on many aspects controlled.  Therefore, one can foresee three very interesting  classes of 
possible applications of harmonic pulses: applications in nonlinear optics in the XUV, that employ the high refocused harmonic intensities,
applications in interferometry, that employ spatial coherence properties,
and applications in attosecond physics, that employ the phase modulation and other phase properties to reduce the duration of harmonic pulses.
In this Chapter we present a somewhat speculative discussion of these three classes of applications, since none of them has been realized so far. 

\subsection{Nonlinear optics in the XUV}

In Chapter 4, we have shown that the good wavefront quality of the harmonic beam could allow us to refocus it quite easily to very small spots of a few $\mu$m in diameter using simply  a spherical mirror. Recently, the Saclay group has  confirmed this prediction
by measuring  the size of the focal spot, and its dependence on the medium characteristics \cite{lederoff}. Even better focusing could be achieved  by using a Schwarzschild objective \cite{schwarz}.

If the short pulse character of the emission is preserved, which implies the use of multilayer mirrors to select and refocus the harmonic beam, unprecedented high intensities could be reached in the XUV region. For instance, if argon is used as generating medium, an optimized number of 10$^{10}$ photons could be generated around 30 eV. Schafer and Kulander \cite{kenpriv} found that the conversion efficiency at the single atom level increases roughly as quickly as the laser pulse duration decreases, resulting in the emission of about the same number of harmonic photons. Using the intense 25 fs infrared lasers currently available, an harmonic pulse duration of less than 10 fs could be obtained (note that the generation of sub-femtosecond pulses is under consideration, see below). The refocused harmonic intensity would thus reach 10$^{14}$ W/cm$^2$, enough to induce multiphoton processes. 

Some attempts have already been done to observe non-resonant two-photon ionization of rare gases \cite{bouh,VanW,koba}. They are for the moment unsuccessfull, except for moderate-order harmonics \cite{koba}, but the rapid progress in short pulse laser technology (repetition rate, pulse duration and energy) will improve the performances of the harmonic sources and should make possible the observation of such processes in the near future.

\subsection{Interferometry with harmonics}

Interferometry with harmonics is particularly interesting because of their high frequency and short duration. Coherent harmonics of high frequencies would pass many interesting media without significant absorbtion. Of particular interest are the dense laser-induced plasmas. Short wavelengths are much less sensitive than visible ones to refraction by the large density gradients involved, and correspond moreover to higher critical densities.  Furthermore, due to their short duration, the harmonics may probe dynamics of many systems in a quasi-instantaneous way. Interferometry with harmonics thus appear as a powerful diagnosis for different medias, and in particular plasmas. The broad harmonic bandwith (correlated to the short pulse duration) makes rather difficult
to  use an amplitude division interferometer (Michelson's type) at the harmonic frequency,  but the good spatial coherence is well suited for wavefront division interferometers (Young's type). 
Two kinds of possible interferometry
 experiments are 
discussed by experimentalists: in one kind,  two correlated harmonic beams are used
(like in \cite{zerne}), out of which one may, and the other does not
pass through the medium to be diagnosed; both beams interfere thereafter.
Realization of such experiment may pose, however, difficulties in achieving sufficient
spatial and temporal overlap of the two harmonic pulses.

The other kind of experiment involves only one harmonic beam which would pass through a region of spatially non-uniform index of refraction,
and produce a self-interference pattern thereafter. Numerous
applications of this last scheme in solid state and plasma diagnostics are currently being considered by experimentalists in Saclay.

Note finally that the harmonic spatial coherence could be used in holographic or phase-contrast imaging. 
In particular, the harmonic generation spectrum has now reached the water window (between the K-edges of carbon at 4.4 nm and oxygen at 2.3 nm \cite{Spielmann,Changprl}). This region is of particular interest for biological applications since it provides the best contrast for hydrated carbonated structures such as living cells.

\subsection{Attosecond physics}

The shortest pulses achieved today (5 fs in the infrared \cite{Baltuska}) are limited by the long period of the radiation. Shorter wavelengths are required for further pulse shortening. High-order harmonics are the most promising way to generate subfemtosecond, i.e. attosecond (as), pulses. Different schemes have been proposed to reach these extremely short durations. The first one deals with the relative phase of the harmonics generated in the plateau region \cite{Farkas,Harris}. If they were emitted in phase, the corresponding temporal profile would consist of a {\it train} of pulses separated by half the laser period and of duration in the attosecond range. There is a clear analogy here with mode-locked lasers. However, early calculations of the single atom response showed that the high harmonics were, in general, not in phase, due to the interference  of various energetically allowed electronic trajectories leading to the harmonic emission. Recently, Antoine {\it et al.} revived this proposal by showing that the propagation in the atomic medium could select one of these trajectories, resulting in the macroscopic emission of a train of $\approx$ 200 attosecond pulses \cite{Antoine}.

In principle, the proposals based on phase locking require  filtering of
the phase locked harmonics from the entire spectrum; in particular they require suppression of the low harmonics from the fall-off region of the spectrum. It is worth mentioning
that trains of attosecond pulses could also be generated in laser-induced plasmas from the surface of a solid target.  Harmonic generation from solid targets
has been  intensively studied in the recent years both experimentally
\cite{co2,wahl2,vonder,norrey}, and theoretically 
\cite{Gibbon,pukhov,lichter}, but only very recently have the temporal aspects of such a generation process been studied \cite{roso1,roso2}.   Using a simple model of the
oscillating
plasma surface, as well as a 1D ``Particle--In--Cell'' code, these authors have shown  that the
reflected signal obtained in the case of normal incidence of the
driving
laser has the form of a train of ultrashort pulses of duration
$\simeq20$ times
shorter than the
optical period (i.e. $\simeq 100$ attoseconds for the case of Ti-
Sapphire
laser).
This is quite surprising because the harmonic spectrum is
monotonically
decreasing and does not
exhibit any plateau, and moreover, the production of
those
pulses does not require any filtering.

Furthermore, Corkum {\it et al.} \cite{Corkum2,Ivanov} proposed a way of generating a {\it single} pulse by using the high sensitivity of harmonic generation in gases to the laser polarization \cite{Budil,Dietrich}. By creating a laser pulse whose polarization is linear only during a short time, close to a laser period, the emission could be limited to this interval. Propagation calculations validated this idea \cite{Antoine2}. In their original proposal, Corkum {\it et al.} suggested to use two cross polarized fundamental beams of slightly different frequencies in order to obtain the desired 
form of the time dependent polarization.  Very recently,
the Lund group has realized this idea in a somewhat different way \cite{wahl}, by using a flat birefringent crystal and a short input pulse  that is chirped in frequency and polarized at 45 degrees relative to the optical axis of the crystal. The input chirped pulse in such a system is split into two orthogonally polarized chirped pulses delayed in time. When the phase difference between the two polarizations is zero, linear polarization is produced. This enables a single ultrashort time window to be selected for efficient harmonic generation. Moreover, two or more ultrashort time windows with controllable delay may be selected.

Another scheme of ultrashort pulse generation has been described in the preceding section, and concerns a single harmonic. The compression of the dynamically induced frequency chirp of an harmonic located at the end of the plateau and generated by a 27 fs laser pulse can result in a pulse of 1.7 fs duration. Moreover, Schafer and Kulander proposed to compress three adjacent harmonics to generate $\approx$ 400 as pulses \cite{kenpriv}. This is made possible by the fact that harmonics located close to the cutoff exhibit approximately the same frequency chirp.

Finally, the last scheme uses very short duration laser pulses to shorten even more the harmonic emission, thanks to the rapid ionization of the medium (less than one cycle). As already mentioned,
with 5 fs infrared pulses it has become recently
possible to generate soft XUV radiation extending to the K-edge of carbon at 4.4 nm (water window) \cite{Spielmann}. Simulations show that the generated pulses could be of sub-500 attosecond duration \cite{Christov2}.

Although the problem of generating attosecond pulses is already difficult, it is even more challenging to consider their detection, and applications. So far, the literature on those subjects hardly exist, but theoretical studies have begun. Corkum and his collaborators \cite{corkum97}
discuss possible applications of subfemtosecond pulses in physical chemistry.
These authors discuss, for instance, Coulomb explosion 
measurements for molecules in strong laser fields; such measurements
allow to measure time-dependent molecular structures. The use of  attosecond pulses for such measurments would allow to "freeze" the nuclear motion for even highly charged molecular ions by their inertia for the duration of the pulse,
and thus achieve  much better temporal resolution than it is possible so far.

An experimental 
detection of the train of attosecond pulses is equivalent to the detection of phase locking. The latter can be studied for instance in the process of two-photon ionization of Helium atoms by a filtered harmonic signal
consisting of 5-10 harmonics. The ionization yield in such a process should depend on the relative phases between the harmonics \cite{agop}.
Maquet and his collaborators in a series of papers has investigated the possibility of determining the relative phases between the harmonics and the fundamental field from the above threshold ionization (ATI) spectra. 
In Refs. \cite{ven1,tai1} ATI in  a
two color field (fundamental plus one high harmonic frequency) was cansidered. The spectra in that case may be strongly affected
 by quantum interferences  between  different 
ionization and electron redistribution 
 channels, and depend on the relative phase between the two fields.  
In further publication  \cite{ven2} the authors considered ATI in a mixed 
multicolor field consiting of the  fundamental and several high harmonic components. The detailed study of photoionisation spectra provides in this
case more, or less direct information on the phase differences between successive harmonics. Comparison with experimental results \cite{agop}
suggests that indeed harmonics generated in rare gases are phase locked.

Another possibility of detection and application of attosecond pulses
would be to develop the time resolved attosecond spectroscopy (TRAS). This could be done in analogy with femtosecond spectroscopy, using a pump and probe attosecond pulses, for example to excite
and then ionize a coherent sub-Rydberg wave packet in an atom. The result of such pump--probe process should depend on the time delay between the pump and the probe.

In the review \cite{phrev} we suggested that TRAS could also be realized by mixing an attosecond pulse train (the probe) with the strong fundamental laser beam (the pump). In particular, we have considered harmonic generation by such multicolored fields, expecting that the resulting harmonic signal would depend on the time delay (modulo one fundamental period) between the train and the pump. Unfortunately, as we already pointed out in \cite{phrev},
the estimates presented in that paper, based on SFA, are too optimistic.
We have checked, using TDSE method, that the dependence of the harmonic signal 
on the time delay in such process is much weaker
 in the experimentally accessible regime of parameters. 
Nevertheless, the scheme allows at least to study temporal interference 
between the train of pulses focused in the medium (probe), and the train of pulses generated in the medium by the fundamental laser (pump). This
interference can be particularly efficient due to phase matching effects, 
such as trajectory selection via propagation  \cite{Antoine}.

Other processes than harmonic generation,
 can  be considered in order to make TRAS with attosecond pulse trains, or separated attosecond pulses possible. In Ref.\cite{tai2} the authors consider
ATI process  by a fundamental laser pulse that leads to the resonant excitation of dynamically Stark-shifted Rydberg states. Short harmonic pulse (of duration much shorter than the fundamental one) is applied during that process, and its 
delay with respect to the front of the fundamental pulse is controlled.
As a result, the highly structured ATI spectra depend on the delay of the
probe,  and provide thus the first 
example of time-resolved spectrocopy of dynamically
induced resonances. The quest for other forms of TRAS remains, in our opinion, one of the most interesting challenges of super-intense laser atom physics.

\section{Conclusion}

We have used a theory of high order harmonic generation by low frequency laser fields in the Strong Field Approximation to study the spatial and temporal coherence properties of the harmonic radiation. We show that the intensity dependence of the atomic dipole phase predicted by this model plays a central role in the way phase matching is achieved in the medium. In particular, it introduces an asymmetry relative to the position of the jet compared to the laser focus. When the laser is focused sufficiently before the jet, phase matching is optimized on axis, leading to very regular Gaussian emission profiles. When it is focused in or after the medium, harmonic generation is prevented on axis but can be efficient off axis resulting in annular profiles. Moreover, the dipole phase induces a big curvature on the generated harmonic phase front, leading to quite diverging harmonic beams. Consequently the virtual source of the emission presents a very small size. Temporal fluctuations of the harmonic phase front can degrade the spatial coherence of the radiation, in particular for a focusing after the jet. However, before the jet, the spatial coherence can be very high indicating that high order harmonics could be a useful coherent source in the XUV.

Induced by the intensity distribution in the laser pulse, the dynamic phase matching results in very regular, narrow temporal profiles for focus positions sufficiently before or after the jet. At intermediate positions, the harmonic profiles are distorted and almost as large as the fundamental. The corresponding spectra are very broad and far from the Fourier transform limit. This is due to the intensity dependence of the dipole phase, which results in a phase modulation of the emitted harmonic field. This phase modulation is all the more important as the peak intensity is high and the harmonic emission is in the plateau region, where the slope of the intensity dependence is the larger. The harmonic beam thus presents a negative chirp, with a rising edge shifted to the blue and a falling edge, to the red. In our conditions, ionization is quite limited and affects mainly the falling edge, and thus the red part of the spectrum. 

The regular phase modulation can be used to control the temporal and spectral properties of the harmonic radiation. Temporally, the chirped pulse could be recompressed by a pair of gratings to very small durations. Spectrally, an appropriately designed chirp on the fundamental can compensate (or increase, depending on its sign) the harmonic dynamic chirp, resulting in a narrow (or extremely broad) spectrum. Finally, we show that cutoff harmonics generated by pulses as short as 27 fs can be described by an adiabatic calculation, the nonadiabatic phenomena being negligible. 

In conclusion, the spatial and temporal coherence properties of high order harmonics are closely related, and depend strongly on the atomic dipole phase. However, the regular behavior of the latter makes it 
possible to control them by using relatively simple means.

These unprecedented coherence properties in the XUV region open the way to a new class of potential applications which are attracting more and more interest. The good wavefront quality, that allows focusing to very small focal spots, together with the short pulse duration could result in the high focused intensities necessary to perform nonlinear optics in the XUV. The high spatial coherence makes it possible to realize XUV interferometry experiments, such as ultrashort plasmas diagnostics. Finally, the harmonic phase characteristics in time could be used to generate attosecond pulses, the shortest ever, opening the way to "attosecond physics".

We acknowledge fruitful discussions with P. Agostini, Th. Auguste, K. Burnett, B. Carr\'e, P. B. Corkum, T. Ditmire, M. D\"orr, M. Gaarde, M. Y. Ivanov, Ch. Joachain, H. C. Kapteyn, K. C. Kulander, A. Maquet, P. Monot, M. M. Murname, M. D. Perry, A. Sanpera, K. J. Schafer, R. Ta\"\i eb, V. V\'eniard, C.-G. Wahlstr\"om, and J. Watson.

\end{document}